\documentclass{article}
\usepackage{arxiv}

\usepackage[utf8]{inputenc} 
\usepackage[T1]{fontenc}    
\usepackage{hyperref}       
\usepackage{url}            
\usepackage{booktabs}       
\usepackage{amsfonts}       
\usepackage{nicefrac}       
\usepackage{microtype}      
\usepackage{lipsum}
\usepackage[version=3]{mhchem} 
\usepackage{mathptmx}
\usepackage{multirow}
\usepackage{hyperref}
\usepackage{braket}
\usepackage{rotating}
\usepackage{color}
\usepackage{graphicx}
\usepackage{amsmath}
\usepackage[square,numbers,compress]{natbib}
\usepackage{notes2bib}
\usepackage{comment}

\newcommand{\fig}{Fig.}
\newcommand{\Fig}{Fig.}
\newcommand{\figref}[1]{\fig~\ref{#1}}
\newcommand{\Figref}[1]{\Fig~\ref{#1}}

\newcommand{\Tabref}[1]{Table~\ref{#1}}

\renewcommand{\eqref}[1]{Eq.~(\ref{#1})}

\newcommand{\secref}[1]{Section~\ref{#1}}

\providecommand{\e}[1]{\ensuremath{\times 10^{#1}}}

\DeclareMathOperator*{\argmin}{argmin}

\title{Fast and Sample-Efficient Interatomic Neural Network Potentials for Molecules and Materials Based on Gaussian Moments}

\author{      Viktor Zaverkin$^1$\\
              University of Stuttgart\\
              Faculty of Chemistry\\
              Institute for Theoretical Chemistry\\
              \And
              David Holzmüller$^1$\\
              University of Stuttgart\\
              Faculty of Mathematics and Physics\\
              Institute for Stochastics and Applications\\
              \And
              Ingo Steinwart\\
              University of Stuttgart\\
              Faculty of Mathematics and Physics\\
              Institute for Stochastics and Applications\\
              \And
              Johannes Kästner$^2$\\
              University of Stuttgart\\
              Faculty of Chemistry\\
              Institute for Theoretical Chemistry\\
              }

\begin{document}
\footnotetext[1]{Both authors contributed equally to this work.}
\footnotetext[2]{E-Mail: \texttt{kaestner@theochem.uni-stuttgart.de}}
\maketitle

\begin{abstract}
Artificial neural networks (NNs) are one of the most frequently used machine learning approaches to construct interatomic potentials and enable efficient large-scale atomistic simulations with almost ab initio accuracy. However, the simultaneous training of NNs on energies and forces, which are a prerequisite for, e.g., molecular dynamics simulations, can be demanding. In this work, we present an improved NN architecture based on the previous GM-NN model [V. Zaverkin and J. Kästner, J. Chem. Theory Comput. 16, 5410-5421 (2020)], which shows an improved prediction accuracy and considerably reduced training times. Moreover, we extend the applicability of Gaussian moment-based interatomic potentials to periodic systems and demonstrate the overall excellent transferability and robustness of the respective models. The fast training by the improved methodology is a pre-requisite for training-heavy workflows such as active learning or learning-on-the-fly.
\end{abstract}

\keywords{Gaussian moments \and Atomistic neural networks \and Computational chemistry \and Computational materials science}

\section{\label{sec:intro} Introduction}

Approximate methods, such as empirical force fields (FFs)~\cite{Hornak06, Vanommeslaeghe10, Halgren96}, are an integral part of modern computational chemistry and materials science. While the application of first-principles methods, such as density functional theory (DFT), to even moderately sized molecular and material systems is computationally very expensive, approximate methods allow for simulations of large systems over long time scales. During the last decades, machine-learned potentials (MLPs)~\cite{Behler2010, Bartok2010, Rupp2012, Bartok2013, Linienfeld2015, Shapeev2016, Khorshidi2016, Schuett2017, Artrith2017, Faber2018, Zhang2018, Zhang2018_2, Kocer2019, Zhang2019, Christensen2020, Zaverkin2020, Schuett2021, Behler2007, Behler2011, Artrith2016, Schuett2017_2, Chmiela2017, Chmiela2018, Gubaev2018, Lubbers2018, Yao2018, Unke2019, Cooper2020, Smith2019, Sivaraman2021} have risen in popularity due to their ability to be as accurate as the respective first principles reference methods, the transferability to arbitrary-sized systems, and the capability of describing bond breaking and bond formation as opposed to empirical FFs~\cite{Mackerell2004}.

Interpolating abilities of neural networks (NNs)~\cite{Hornik1991} promoted their broad application in computational chemistry and materials science. NNs were initially applied to represent potential energy surfaces (PESs) of small atomistic systems~\cite{Blank1995, Lorenz2004} and were later extended to high-dimensional systems~\cite{Behler2007}. Once trained, the computational cost of MLPs based on NNs does not scale with the number of data points used for training as opposed to kernel-based models~\cite{Bartok2010, Rupp2012, Bartok2013, Linienfeld2015}. Therefore, training sets can be as large as necessary to achieve the desired interpolation accuracy for applications in atomistic simulations, such as Monte Carlo (MC) sampling or molecular dynamics (MD).

A central ingredient for the construction of robust and accurate MLPs is a carefully chosen molecular representation. A wide variety of such representations have been proposed in the literature~\cite{Behler2010, Bartok2010, Rupp2012, Bartok2013, Linienfeld2015, Shapeev2016, Khorshidi2016, Schuett2017, Artrith2017, Faber2018, Zhang2018, Zhang2018_2, Kocer2019, Zhang2019, Christensen2020, Zaverkin2020, Schuett2021}.
In our previous work~\cite{Zaverkin2020}, we have shown that geometric moments can be successfully used for this purpose. Combined with radial basis functions and used with neural networks, we arrived at Gaussian moment neural networks (GM-NN)~\cite{Zaverkin2020}. We have proposed an NN architecture that allowed us to use a single NN for all atomic species, in contrast to using an individual NN for each species as frequently necessary previously~\cite{Behler2007, Artrith2016, Artrith2017}. This improved the times needed for an individual evaluation of energies and atomic forces.

In this work, we present an improved GM-NN (iGM-NN) architecture for molecular and materials systems, which allows for very fast training, 10--20 times faster compared to the previous version, and shows improved prediction accuracy as well as excellent transferability to configurations not seen during training. Besides the improved NN architecture, including more flexible radial functions used for the construction of the Gaussian-moment representation, we extend the latter to the application on periodic systems.

While our improved NN allows for convenient training on a given data set, it is especially useful when building the training data set on-the-fly, e.g., during a molecular dynamics (MD) simulation, or offline, drawing new training structures from an unlabeled data set. These workflows are referred to as learning-on-the-fly or active learning~\cite{Settles2009, Li15, Podryabinkin17, Smith18, Zhang19, Gastegger17, Gubaev2018, Zaverkin2021, Janet19, Vandermause20}, respectively. They reduce the number of required ab-initio calculations, but require frequent re-training of the model from updated training sets. The application of state-of-the-art machine learning models based on artificial NNs in such training-heavy workflows is somewhat hindered by their training time, which often ranges from several hours to several days~\cite{Unke2019, Schuett2019, Zaverkin2020}. We expect that our short training times will render active learning and learning-on-the-fly approaches much more attractive. Moreover, without access to GPUs, one can still train iGM-NN quickly on a CPU-only system.

To assess the quality of MLPs based on the improved methodology, we thoroughly benchmark the predictive accuracy as well as inference and training times on established molecular data sets from the literature, QM9~\cite{Ruddigkeit2012, Ramakrishnan2014} and MD17~\cite{Chmiela2017, Schuett2017_2, Chmiela2018}. Finally, we studied two solid-state systems, \ce{TiO2}~\cite{Artrith2016} and \ce{Li8Mo2Ni7Ti7O32}~\cite{Cooper2020}, to investigate the applicability of iGM-NN potentials to periodic systems as well as to assess their robustness and transferability during real-time atomistic simulations. 

We provide an open-source implementation at \href{https://gitlab.com/zaverkin\_v/gmnn}{gitlab.com/zaverkin\_v/gmnn} and \href{https://doi.org/10.18419/darus-2136}{doi.org/10.18419/darus-2136}.

\section{\label{sec:theory} Theory}
This section first introduces the representation used to describe the local atomic environments of atoms in molecular and solid-state systems throughout this work and the changes compared to our previous work~\cite{Zaverkin2020}. Second, we describe the architecture of the feed-forward neural network used to learn the non-linear map between physicochemical properties and the representation as well as its training.

\subsection{\label{sec:representation} Representation}
In this work, we denote an atomic structure as $S^{\left(k\right)} = \{\mathbf{r}_i, Z_i\}_{i=1}^{N_\mathrm{at}^{\left(k\right)}}$, where $\mathbf{r}_i \in \mathbb{R}^3$ are the Cartesian coordinates of atom $i$ and $Z_i \in \mathbb{N}$ encodes its species (e.g., $Z_i=1$ for a H atom).
A potential energy surface (PES) is a function that maps an atomic structure to a scalar energy, i.e. $f: S^{\left(k\right)} \mapsto E \in \mathbb{R}$. The purpose of molecular machine learning (ML) approaches is to learn this mapping without solving the electronic Schr{\"o}dinger equation.  

A machine-learned PES has to satisfy several symmetries. These are the invariance with respect to global rotations, translations, and reflections of a molecular structure, as well as with respect to the exchange of atoms of the same atomic species. Additionally, it should allow the generalization to larger structures. The latter requirement is satisfied by decomposing the total energy of a system into a sum of ``auxiliary'' atomic energies~\cite{Behler2007}

\begin{equation}
    \hat{E} \left(S^{\left(k\right)}, \boldsymbol{\theta}\right) \approx \sum_{i=1}^{N_\mathrm{at}} \hat{E}_{i}\left( \mathbf{G}_i, \boldsymbol{\theta} \right)~,
\end{equation}
where the invariances are encoded in the local representation $\mathbf{G}_i$ and $\boldsymbol{\theta}$ are the trainable parameters of the ML model. Making the representation dependent on the atomic species, i.e. $\mathbf{G}_i = \mathbf{G}_i\left(\mathbf{r}_i, Z_i,  \{\mathbf{r}_j, Z_j\}_{j=1}^{N_\mathrm{c}}\right)$ where $N_\mathrm{c}$ is the number of atoms in the cutoff sphere, it is possible to train a single neural network (NN) for all species in the training set~\cite{Zaverkin2020}, different to, e.g., Ref.~\citenum{Behler2007}. Note that in our approach the local representation is trainable, too.

\begin{figure}[t]
\centering
\includegraphics[width=6cm]{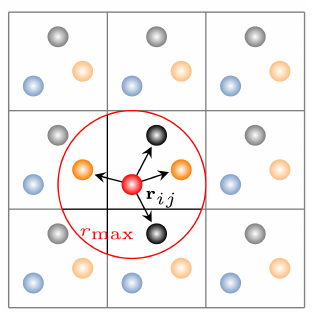}
\caption{Two-dimensional schematic illustrating the local environment of an atom.}
\label{fig:pbc_neighbors}
\end{figure}

One of the main challenges is to find an appropriate molecular representation that introduces all required symmetries into the ML model. Among the suite of existing invariant molecular representations, we build upon the trainable Gaussian moments (GM) that we introduced recently~\cite{Zaverkin2020}. In this approach, the atomic distance vectors $\mathbf{r}_{ij} = \mathbf{r}_i - \mathbf{r}_j$ are split into their radial and angular components, i.e. $r_{ij} = \lVert\mathbf{r}_{ij}\rVert_2$ and $\hat{\mathbf{r}}_{ij} = \mathbf{r}_{ij} / r_{ij}$ respectively. Note that for a periodic system one has to include the distance vectors to periodic images of atoms in the local neighborhood, see \figref{fig:pbc_neighbors}. A tensor-valued function of the local atomic environment can then be written as
\begin{equation}
\label{eq:psi}
    \begin{split}
        \boldsymbol{\Psi}_{i, L, s} = \sum_{j \neq i} R_{Z_i, Z_j, s}\left(r_{ij}, \boldsymbol{\beta}\right)  \hat{\mathbf{r}}_{ij}^{\otimes L}~,
    \end{split}
\end{equation}
where $\hat{\mathbf{r}}_{ij}^{\otimes L} = \hat{\mathbf{r}}_{ij} \otimes \cdots \otimes \hat{\mathbf{r}}_{ij}$ is the $L$-fold tensor product of the angular components and $R_{Z_i, Z_j, s}\left(r_{ij}, \boldsymbol{\beta}\right)$ are nonlinear radial functions with trainable parameters $\boldsymbol{\beta}$. Note, that in difference to Ref.~\citenum{Zaverkin2020} we extend the trainable parameter by an index $s$ from 1 to $N_\mathrm{Basis}$, i.e. $\beta_{Z_i, Z_j, s, s^\prime}$ with $s^\prime$ being index from 1 to $N_\mathrm{Gauss}$. The radial function has, similar to the one presented in Ref.~\citenum{Gubaev2018}, the form
\begin{equation}
    \label{eq:radial_function}
    R_{Z_i, Z_j, s}\left(r_{ij}, \boldsymbol{\beta}\right) = \frac{1}{\sqrt{N_\mathrm{Gauss}}}\sum_{s^\prime=1}^{N_\mathrm{Gauss}}\beta_{Z_i, Z_j, s, s^\prime} \Phi_{s^\prime}\left(r_{ij}\right)~, 
\end{equation}
where $N_\mathrm{Gauss}$ is the number of Gaussian functions $\Phi_{s^\prime}\left(r_{ij}\right)$ placed along the radial coordinate and the factor $1/\sqrt{N_\mathrm{Gauss}}$ influences the effective learning rate during optimization inspired by the NTK parameterization~\cite{Jacot2018}. The radial function 
\begin{equation}
    \Phi_{s^\prime}(r_{ij}) = \left(\frac{2 N_\mathrm{Gauss}}{\pi r_\mathrm{max}^2}\right)^{1/4} e^{-N_\mathrm{Gauss}^2 (r_{ij} - \gamma_{s^\prime})^2 / r_\mathrm{max}^2}f_\mathrm{cut}(r_{ij})
\end{equation}
is centered at $\gamma_{s^\prime} = r_\mathrm{min} + \frac{s^\prime-1}{N_\mathrm{Gauss}-1} (r_\mathrm{max} - r_\mathrm{min})$ and re-scaled by the cosine cutoff function~\cite{Behler2007} $f_\mathrm{cut}(r)$ with a cutoff radius $r_\mathrm{max} > r_\mathrm{min} = 0.5$~\AA{}.

We initialize the trainable coefficients $\beta_{Z_i, Z_j, s, s'}$ independently uniformly on the interval $[-1, 1]$, similar to Ref.~\citenum{Zaverkin2020}. However, we have found that on the QM9 data set~\cite{Ruddigkeit2012, Ramakrishnan2014}, better accuracy can be achieved by initializing $\beta_{Z_i, Z_j, s, s'} = \delta_{s, s'}$. We assume that the reason for this is that the distance $r_{ij}$ between atoms already contains some information on the species $Z_i$ and $Z_j$ if the respective data set contains only structures in equilibrium, like the QM9 data set.

The tensors $\boldsymbol{\Psi}_{i, L, s}$ are invariant with respect to translations and to permutations of atoms of the same atomic species. A rotationally invariant representation is obtained by performing full contractions
\begin{equation}
\label{eq:contractions}
\begin{split}
    & G_{i, s_1}^{(1)} = \boldsymbol{\Psi}_{i,0,s_1}~,\\
    & G_{i, s_1, s_2}^{(2)} = \left(\boldsymbol{\Psi}_{i,1,s_1}\right)_a \left(\boldsymbol{\Psi}_{i,1,s_2}\right)_a~,\\
    & G_{i, s_1, s_2}^{(3)} = \left(\boldsymbol{\Psi}_{i,2,s_1}\right)_{a,b}\left(\boldsymbol{\Psi}_{i,2,s_2}\right)_{a,b}~,\\
    & G_{i, s_1, s_2}^{(4)} =  \left(\boldsymbol{\Psi}_{i,3,s_1}\right)_{a,b,c}\left(\boldsymbol{\Psi}_{i,3,s_2}\right)_{a,b,c}~,\\
    & G_{i, s_1, s_2, s_3}^{(5)} = \left(\boldsymbol{\Psi}_{i,1,s_1}\right)_{a}  \left(\boldsymbol{\Psi}_{i,1,s_2}\right)_{b}\left(\boldsymbol{\Psi}_{i,2,s_3}\right)_{a,b}~,\\
    & G_{i, s_1, s_2, s_3}^{(6)} =  \left(\boldsymbol{\Psi}_{i,2,s_1}\right)_{a,b}  \left(\boldsymbol{\Psi}_{i,2,s_2}\right)_{a,c}  \left(\boldsymbol{\Psi}_{i,2,s_3}\right)_{b,c}~,\\
    & G_{i, s_1, s_2, s_3}^{(7)} =  \left(\boldsymbol{\Psi}_{i,1,s_1}\right)_{a}  \left(\boldsymbol{\Psi}_{i,3,s_2}\right)_{a,b,c}  \left(\boldsymbol{\Psi}_{i,2,s_3}\right)_{b,c}~,\\
    & G_{i, s_1, s_2, s_3}^{(8)} =  \left(\boldsymbol{\Psi}_{i,3,s_1}\right)_{a,b,c}  \left(\boldsymbol{\Psi}_{i,3,s_2}\right)_{a,b,d}  \left(\boldsymbol{\Psi}_{i,2,s_3}\right)_{c,d}~,
\end{split}
\end{equation}
where we use Einstein notation, i.e., the right-hand sides are summed over $a, b, c, d \in \{1, 2, 3\}$. 

By construction, the tensors $\mathbf{G}^{(k)}$ in \eqref{eq:contractions} have certain symmetries. In order to avoid including duplicate elements into the descriptor, in Ref.~\citenum{Zaverkin2020} we used upper triangular elements, i.e., those with indices satisfying $s_1 \leq s_2 \leq s_3$. However, tensors $\mathbf{G}^{(5)}$ and $\mathbf{G}^{(8)}$ posses symmetry only with respect to permutation of indices $s_1$ and $s_2$ whereas $\mathbf{G}^{(7)}$ does not show any symmetry to index permutation. Therefore, using upper triangular elements may lead to the results which depend on the specific order of contractions presented in \eqref{eq:contractions}. To deal with this shortage we now use all elements of $\mathbf{G}^{(7)}$ and only elements with $s_1 \leq s_2$ for $\mathbf{G}^{(5)}$ and $\mathbf{G}^{(8)}$. For $N_\mathrm{Basis}=7$, this yields $N_\mathrm{Feature} = 910$ invariant descriptors instead of $N_\mathrm{Feature} = 427$ used in Ref.~\citenum{Zaverkin2020} at almost no computational overhead since in a practical implementation whole tensors $\mathbf{G}^{(k)}$ are computed. Additionally, this allows us to reduce the number of basis functions to, e.g., $N_\mathrm{Basis}=5$ corresponding to $N_\mathrm{Feature} = 360$ invariant features, which reduces the overall computational cost.

\subsection{\label{sec:machine_learning} Machine Learning}

\subsubsection{\label{sec:feature_norm} Normalization of Input Features}

The normalization of feature vectors prior to training a neural network is known to improve the performance of the latter and, therefore, is usually applied to neural networks. For an architecture that uses trainable representation based on Gaussian moments (GM) we have found it to be redundant~\cite{Zaverkin2020} except for data sets such as QM9~\cite{Ruddigkeit2012, Ramakrishnan2014}, a quantum chemistry data set that covers a wide range of chemical space. For the latter, we have found that normalizing the input features leads to a considerable improvement of the predictive accuracy. This is likely due to the specific initialisation of $\beta_{Z_i, Z_j, s, s'}$ for QM9, see \secref{sec:representation}, which results in non-negative radial functions $R_{Z_i, Z_j, s}\left(r_{ij}, \boldsymbol{\beta}\right)$ at initialisation. Therefore, we describe here a way to center and standardize the trainable local representation inspired by the Batch Normalization technique~\cite{Ioffe2015}. First, during training, the mean and the standard deviations over the current minibatch $b$ containing $N_\mathrm{Struct}$ structures are calculated as
\begin{equation}
    \begin{split}
        \boldsymbol{\mu}_b &= \frac{\sum_{k=1}^{N_\mathrm{Struct}^{(b)}} \sum_{i=1}^{N_\mathrm{at}^{(k)}} \mathbf{G}_i}{\sum_{k=1}^{N_\mathrm{Struct}^{(b)}} \sum_{i=1}^{N_\mathrm{at}^{(k)}} 1}~, \\ 
        \sigma_b &= \frac{\sum_{k=1}^{N_\mathrm{Struct}^{(b)}} \sum_{i=1}^{N_\mathrm{at}^{(k)}} \lVert\mathbf{G}_i - \boldsymbol{\mu}_b\rVert_2^2}{\sum_{k=1}^{N_\mathrm{Struct}^{(b)}} \sum_{i=1}^{N_\mathrm{at}^{(k)}} N_\mathrm{Feature}}~.
    \end{split}
\end{equation}

The running statistics are computed as an exponentially moving average over the previous batches, i.e. as
\begin{equation}
    \begin{split}
    \bar{\boldsymbol{\mu}}_b &= \gamma \bar{\boldsymbol{\mu}}_{b-1} + (1-\gamma) \boldsymbol{\mu}_b~, \\
    \bar{\sigma}_b &= \gamma  \bar{\sigma}_{b-1} + (1-\gamma) \sigma_b,
    \end{split}
\end{equation}
where $\bar{\boldsymbol{\mu}}_{0}$ and $\bar{\sigma}_0$ are initialized to zeros. The final values for the scale and shift of feature vector are re-scaled by a factor $s_b = 1 / (1 - t_b + \epsilon)$ with $\epsilon = 10^{-8}$ being a small number used to avoid division by zero and $t_b = \gamma \cdot t_{b-1}$ with $t_{0} = 1$. The factor $1/(1-t_b)$ is used to correct for the bias introduced by initializing $\bar{\boldsymbol{\mu}}_{0}$ and $\bar{\sigma}_0$ to zeros. We obtained good results by setting the momentum as $\gamma = 0.5 ^ {1/N_\mathrm{Batch}}$, where $N_\mathrm{Batch}$ is the number of batches per epoch, such that the information from one epoch ago is decayed by a factor of 0.5. The normalized feature vector reads
\begin{equation}
    \tilde{\mathbf{G}}_i = c_\mathrm{Scale} \frac{\mathbf{G}_i -  s_b\bar{\boldsymbol{\mu}}_b}{\sqrt{s_b\bar{\sigma}_b + \epsilon}}~,
\end{equation}
where $b$ is the index of the last training batch. Different from standard Batch Normalisation, we use the averaged statistics also during training and compute the variance over all feature vectors in the batch instead of normalizing all features individually. The former reduces the noisy nature of single-batch statistics while the latter preserves the relative importance of the features. No trainable scale and shift parameters for the normalized feature vector are used since the normalization layer is followed by a fully connected layer, which can already learn to scale and shift the features. Instead of that, the normalized feature vector is re-scaled by a constant $c_\mathrm{Scale} = 0.4$, the benefits of which might be explained by the detrimental effects of a large initial neural network function~\cite{Lee2020, Chizat2019, Nonnenmacher2020}.

\subsubsection{\label{sec:architecture} Network Architecture} 

Artificial neural networks (NN) have been proven to be capable of approximating any non-linear function relationship~\cite{Hornik1991}. Therefore, they are a perfect candidate for learning the map between the structure and the respective physicochemical property. In this work, we use a fully-connected feed-forward neural network consisting of two hidden layers of the following functional form
\begin{equation}
\begin{split}
    \hat{y}_i &= 0.1 \cdot \mathbf{b}^{(3)} + \frac{1}{\sqrt{d_2}} \mathbf{W}^{(3)} \phi\left(0.1 \cdot \mathbf{b}^{(2)} + \right. \\ & \quad \left. \frac{1}{\sqrt{d_1}} \mathbf{W}^{(2)} \phi \left(0.1 \cdot \mathbf{b}^{(1)} + \frac{1}{\sqrt{d_0}} \mathbf{W}^{(1)} \mathbf{G}_i\right)\right)~,
\end{split}
\end{equation}
where $\mathbf{W}^{(l)}$ and $\mathbf{b}^{(l)}$ are weights and biases of the respective layer $l$. The parameters $0.1$ and $1/\sqrt{d_l}$ correspond to the so-called NTK parameterisation~\cite{Jacot2018}, which is a theoretically motivated parameterisation for fully-connected layers. We initialize weights of the fully-connected part by picking the respective entries from a normal distribution with zero mean and unit variance. The trainable bias vectors are initialized to zero.

Our network parameterization and initialization are motivated by two goals. First, the variance of the neurons should be on the order of one in all network layers and it should be approximately independent of the layer width $d_l$. This ensures that reasonable regions of the activation function are used and that the initialization does not have to be modified when changing the layer widths. Second, the amount that a trainable parameter needs to change during training should be on the order of one. This ensures that the updates performed by the Adam optimizer, which essentially uses scaled normalized gradients, roughly train all parameters equally fast. In contrast to the Kaiming initialization~\cite{He2015}, which has been motivated by the first goal, the NTK parameterization additionally satisfies our second goal since the weights are initialized with unit variance.

As an activation function we use the Swish/SiLU activation function~\cite{Hendrycks2016, Elfwing2018, Ramachandran2018} $\phi(x) = \alpha x/\left(1 + \exp\left(-x\right)\right)$ multiplied by a scalar $\alpha$, instead of the softplus function used in Ref.~\citenum{Zaverkin2020}. We choose $\alpha \approx 1.6765$ such that $\mathbb{E}_{x \sim \mathcal{N}\left(0, 1\right)} \phi(x)^2 = 1$, i.e., the activation function preserves the second moment if the input is standard Gaussian~\cite{Klambauer2017, Arora2019, Lu2020}.

Throughout this work we used two different sizes of the feature vector with $d_0 = 910$ and $d_0 = 360$, obtained using $N_\mathrm{Gauss} = 7$ and $N_\mathrm{Gauss} = 5$ respectively. The computed local molecular representation $\mathbf{G}_i$ passes through two hidden layers with $d_1 = d_2 = 512$ hidden neurons. The output layer has a single $d_3 = 1$ output neuron since we predict a scalar energy.

\subsubsection{\label{sec:scale_shift} Scaling and Shifting the Atomic Energy}

In order to aid the training process, the output of the neural network can be scaled and shifted by the standard deviation $\sigma$ and the mean $\mu$ of the per-atom average of the reference energies in the training set. In Ref.~\citenum{Zaverkin2020} we have shown that the convergence of the model can be improved by making these parameters trainable as well as dependent on the atomic species, i.e. $\sigma_{Z_i}$ and $\mu_{Z_i}$. Here we improve on this idea by a proper initialization of atomic shifts $\mu_{Z_i}$ and scaling parameters $\sigma_{Z_i}$.

We have found that on data sets with different elemental compositions of structures, a better species-dependent initialization for the shifts $\mu_{Z_i}$ can be obtained using linear regression. Specifically, we solve the linear regression problem
\begin{equation}
\label{eq:lin_reg_init}
\begin{split}
    & \hat{\boldsymbol{\delta}} = \argmin_{\boldsymbol{\delta} \in \mathbb{R}^{N_Z}} \sum_{k=1}^{N_\mathrm{Train}} \left(E\left(S_k\right) - \bar{E}\left(S_k\right) - \boldsymbol{\phi}\left(S_k\right)^T \boldsymbol{\delta}\right)^2 + \lambda \lVert\boldsymbol{\delta}\rVert_2^2~, \\
    & \boldsymbol{\phi}\left(S_k\right) = \left(\begin{array}{c}
    \# \{i: Z_i = 1\} \\
    \# \{i: Z_i = 2\} \\
    \vdots \\
    \# \{i: Z_i = N_Z\} \\
    \end{array}\right)~,
\end{split}
\end{equation}
where $\bar{E}\left(S_k\right) = N_\mathrm{at}^{(k)} \mu$ is the structure's mean energy, $N_Z$ is the number of species, $\lambda = 1$ is a regularization parameter, and $\boldsymbol{\delta}$ are the learned species-dependent differences to the mean atomic energy.
Here, only the difference is regularized, which makes this method invariant to shifts in the total energy. This linear regression problem can be solved in a matter of seconds and scales linearly with the training set size.

Using the output $\hat{y}_i$ of the fully-connected layers, we compute the predicted atomic energy as
\begin{equation}
\label{eq:scale_shift}
\hat{E}_{i}\left( \mathbf{G}_i, \boldsymbol{\theta} \right) = c \cdot (\sigma_{Z_i} \hat{y}_i + \mu_{Z_i})~,
\end{equation}
where $\sigma_{Z_i}$ and  $\mu_{Z_i}$ are trainable parameters, initialized to $1$ and to $(\mu + \hat{\delta}_{Z_i}) / c $, respectively. We set the constant $c$ to be the root-mean-squared error (RMSE) per atom of the regression solution from \eqref{eq:lin_reg_init}, i.e.
\begin{equation}
    c = \sqrt{\frac{1}{N_\mathrm{total}}\sum_{k=1}^{N_\mathrm{Train}}\frac{\left(\bar{E}\left(S_k\right) - E\left(S_k\right)\right)^2}{N_\mathrm{at}^{(k)}}}~,
\end{equation}
where $N_\mathrm{total} = \sum_{k=1}^{N_\mathrm{Train}} N_\mathrm{at}^{(k)} $ is the total number of atoms in the training set. The introduction of the constant parameter $c$ is motivated, like our use of the NTK parameterisation, to achieve uniform learning speed across layers and data sets.

\subsubsection{\label{sec:training} Loss Function and Training}

In this work, we are interested in predicting total energies as well as atomic forces. Therefore, we minimize the following combined loss function
\begin{equation}
    \label{eq:loss}
    \begin{split}
        \mathcal{L}\left(\boldsymbol{\theta}\right) = \sum_{k=1}^{N_\mathrm{Train}} & \Bigg[\lambda_E \lVert E_k^\mathrm{ref} - \hat{E}(S_k, \boldsymbol{\theta})\rVert_2^2 +  \\ & \quad  \frac{\lambda_F}{3N_\mathrm{at}^{(k)}} \sum_{i=1}^{N_\mathrm{at}^{(k)}} \lVert \mathbf{F}_{i,k}^\mathrm{ref} - \hat{\mathbf{F}}_i\left(S_k, \boldsymbol{\theta}\right)\rVert_2^2\Bigg]~,
    \end{split}
\end{equation}
to optimize the respective parameters of the trainable representation, fully-connected neural network part as well as the parameters which scale and shift the output of the neural network. The reference values for the energy and atomic force are denoted by $E_k^\mathrm{ref}$ and $\mathbf{F}_{i,k}^\mathrm{ref}$, respectively. The atomic force for an atom $i$ is calculated by taking the partial derivative of the total energy with respect to the respective atomic position, i.e. $\hat{\mathbf{F}}_i\left(S_k, \boldsymbol{\theta}\right) = -\nabla_{\mathbf{r}_i} \hat{E}\left(S_k, \boldsymbol{\theta}\right)$.

Atomic forces containing $3N_\mathrm{at}$ scalars provide much more information about a molecular structure compared to total energies. Moreover, atomic forces alone determine the dynamics of a chemical system and, therefore, play a crucial role during atomistic simulations such as molecular dynamics simulations. For this purpose it is usual to weight the force error by a larger factor $\lambda_F > \lambda_E$ compared to total energies. In Ref.~\citenum{Zaverkin2020} we have found empirically parameters $\lambda_E = 1$~au and $\lambda_F = 100$~au~\AA$^2$ which already lead to improved predictive accuracy. In this work, we could improve on that by making the force weight dependent on $N_\mathrm{at}$, i.e. $\lambda_F = 12 N_\mathrm{at}$~au~\AA$^2$. For data sets that contain only total energies, like QM9, we use only the energy loss, i.e. $\lambda_F = 0$~au~\AA$^2$.

To minimize the loss function in \eqref{eq:loss} the Adam optimizer~\cite{Adam2015} with hyper-parameters $\beta_1 = 0.9$, $\beta_2 = 0.999$, $\epsilon = 10^{-7}$, and a mini-batch of 32 structures is employed.
Moreover, we allow for layer-wise learning rates which decay linearly to zero by multiplying them with $(1-r)$ where $r = \mathrm{step}/\mathrm{max\_step}$. For all data sets except QM9, we use an initial learning rate of $0.03$ for the parameters of the fully connected layers, $0.02$ for the parameters of the trainable GM representation, $0.05$ and $0.001$ for the shift and scale parameters of atomic energies, respectively. To prevent overfitting on the QM9 data set we use lower learning rates of 0.005, 0.0025, 0.05, and 0.001, respectively. This might be explained by noisier stochastic gradients due to the lack of force labels.

The model is trained for 500 epochs for the QM9 data set containing only total energies as reference values and for 1000 epochs for all other data sets used in \secref{sec:results}. Overfitting was prevented using the early stopping technique~\cite{Prechelt2012}. After each epoch, the mean absolute errors (MAE) of energies and forces were evaluated on the validation set. After training, the model with the minimal sum of energy and force MAEs on the validation set was selected for further application on the test sets. While these hyper-parameters worked reasonably well on the selected benchmarks for us, we want to emphasize that other trade-offs between training speed and accuracy can be achieved by increasing/decreasing the number of epochs and simultaneously decreasing / increasing initial learning rates.

\section{\label{sec:datasets} Data sets}

This section contains a brief description of the data sets used to benchmark the iGM-NN models in \secref{sec:results} on molecular and materials systems.

\subsection{\label{sec:data_qm9} QM9}

The QM9 data set~\cite{Ruddigkeit2012, Ramakrishnan2014} is a widely used benchmark for the prediction of several properties of molecules in equilibrium. Thus, all forces vanish and only energies are used to train the model. The QM9 data set consists of 133,885 neutral, closed-shell organic molecules with up to 9 heavy atoms (C, O, N, F) and a varying number of hydrogen (H) atoms. However, since 3054 molecules from the original QM9 data set failed a consistency test~\cite{Ramakrishnan2014}, we used only the remaining $130,831$ structures in the experiments presented in \secref{sec:results_qm9}. For this data set, a cutoff radius of $r_\mathrm{max} = 3.0$~\AA{} was chosen, similar to Ref.~\citenum{Zaverkin2020}.

\subsection{\label{sec:data_md17} MD17}

The MD17 data set~\cite{Chmiela2017, Schuett2017_2, Chmiela2018} is a collection of structures, energies, and atomic forces of eight small organic molecules obtained from ab-initio molecular dynamics (AIMD). For each molecule, a large variety of conformations is covered. The data set size varies from 150,000 to almost 1,000,000 conformations. It covers energy differences from $20$ to $48$ kcal/mol. Force components range from $266$ to $570$ kcal/mol/\AA. We have chosen a cutoff radius of $r_\mathrm{max} = 4.0$~\AA{} for the MD17 data set, similar to Ref.~\citenum{Zaverkin2020}. Note that we excluded the data set for benzene molecule from experiments since in Ref.~\citenum{Christensen2020} the respective energies were found to be noisy.

\subsection{\label{sec:data_tio2} Bulk \ce{TiO2}}

The \ce{TiO2} bulk data set~\cite{Artrith2016} contains structures, energies, and atomic forces obtained from density functional theory (DFT) calculations using the PBE exchange-correlation functional~\cite{Perdew1996} as implemented in PWSCF of the Quantum ESPRESSO package~\cite{Giannozzi2009}. The data set contains distorted rutile, anatase, and brookite structures as well as the configurations sampled from short molecular dynamics simulations. In addition, supercell structures with oxygen vacancies are included. In total, the \ce{TiO2} data set contains 7815 structures ranging in the size from 6 to 95 atoms. We used a cutoff of $r_\mathrm{max} = 6.5$~\AA{} for training the iGM-NN model in \secref{sec:results_tio2}.

\subsection{\label{sec:data_lmnto} Quaternary Metal Oxide}

To test the applicability of the iGM-NN approach to materials with a more complex chemical composition we used the Li-Mo-Ni-Ti oxide (LMNTO) data set presented in Ref.~\citenum{Cooper2020}. The reference energies and atomic forces are extracted from a 50~ps long AIMD simulation at 400~K. The AIMD simulations of the LMNTO system employed the strongly constrained and appropriately normed (SCAN) semilocal density functional~\cite{Sun2015}. The LMNTO data set contains 2616 periodic structures with 56 atoms each (\ce{Li8Mo2Ni7Ti7O32}).
The data set is available free-of-charge from Ref.~\citenum{Cooper2020_2}. We employed a cutoff radius of $r_\mathrm{max} = 6.5$~\AA{} for training the iGM-NN model in \secref{sec:results_lmnto}.

\section{\label{sec:results} Results}

In this section, we apply the improved Gaussian moment neural network (iGM-NN) architecture to the molecular and materials data sets presented in \secref{sec:datasets}.

\subsection{\label{sec:results_qm9} Results for QM9}

\begin{figure}
    \centering
    \includegraphics[width=8cm]{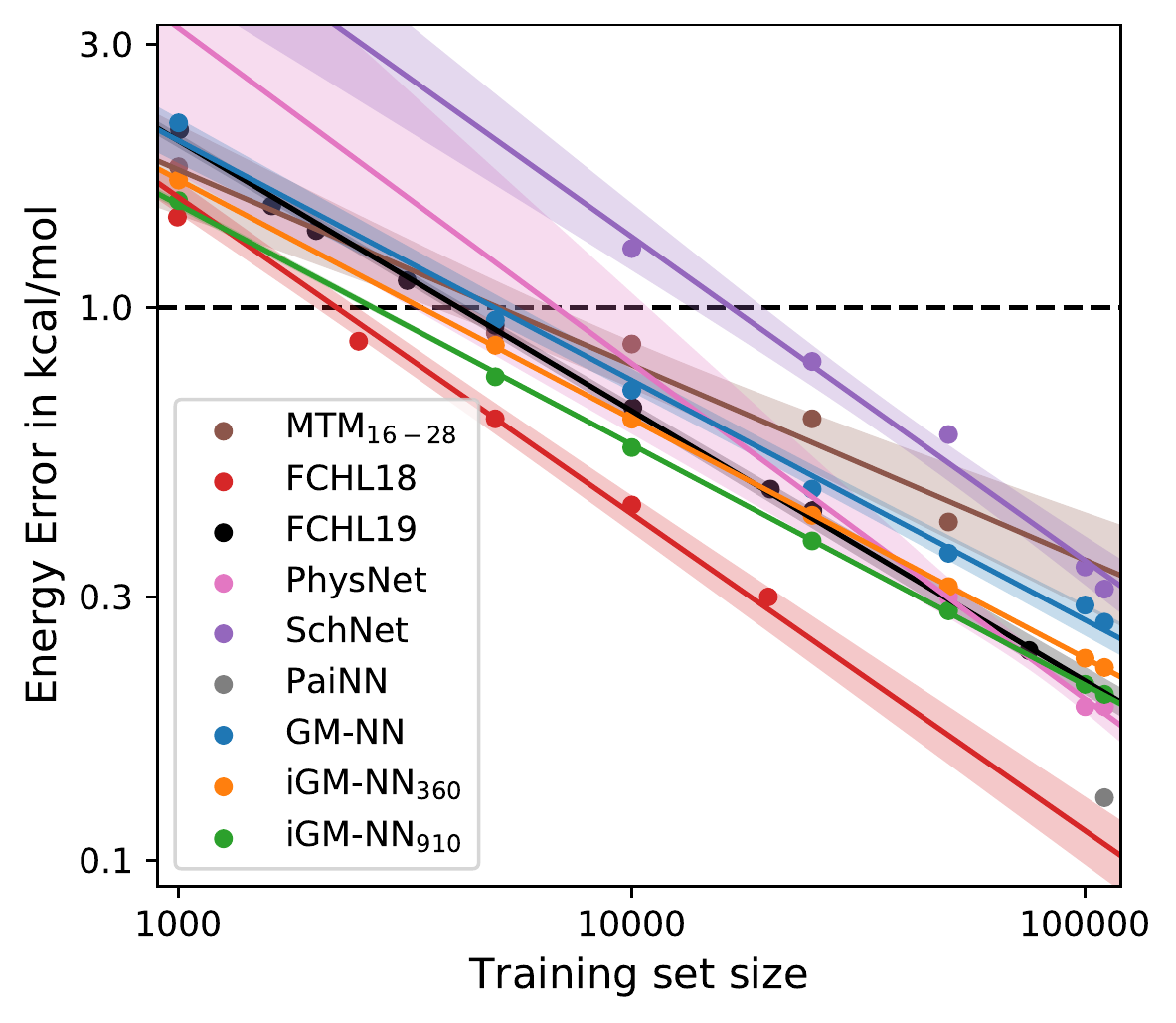}
    \caption{Learning curves for the QM9 data set. The mean absolute error (MAE) of atomization energy is plotted against the training set size. Linear fits are displayed for clarity and shaded areas denote the 95~\% confidence intervals for linear regression. The dashed black line represents the desired accuracy of 1 kcal/mol.}
    \label{fig:qm9_accuracy}
\end{figure}

In \figref{fig:qm9_accuracy}, we compare the predictive accuracy of a number of well-established models for the atomization energy of molecules in the QM9 data set. SchNet~\cite{Schuett2017_2,Schuett2017} and PhysNet~\cite{Unke2019} are message-passing neural networks (MPNNs) with distance-based interactions. PaiNN~\cite{Schuett2021} is a recently proposed rotationally equivariant message-passing neural network architecture. Besides neural network-based approaches, we compare with the kernel-based model FCHL18~\cite{Faber2018} and its faster successor FCHL19~\cite{Christensen2020} as well as with the linear model MTM~\cite{Gubaev2018}. Note that we compare mainly with FCHL19 and PaiNN since these have state-of-the-art accuracy and are also considered to be relatively fast. Finally, we compare the improved Gaussian moment neural network (iGM-NN) model with the previous Gaussian moment neural network (GM-NN) architecture~\cite{Zaverkin2020}.

For the QM9 data set, we find the performance of the iGM-NN model to be among the models with the lowest out-of-sample MAE of atomization energy predictions. Compared to the GM-NN model the MAE at 25000 training samples is 0.38 kcal/mol, 0.42 kcal/mol, and 0.47 kcal/mol for iGM-NN$_{910}$, iGM-NN$_{360}$, and GM-NN respectively. The best performing model, FCHL18, predicts the atomization energy with an MAE of 0.30 kcal/mol and the FCHL19 model with an MAE of 0.47 kcal/mol, both trained on 20000 samples. Beyond that, the iGM-NN model shows a considerable improvement over GM-NN in the predictive accuracy when trained on 1000 samples and shows a comparable accuracy to the relatively slow FCHL18 method. We obtained an MAE of 1.56 kcal/mol, 1.70 kcal/mol, and 2.19 kcal/mol for iGM-NN$_{910}$, iGM-NN$_{360}$, and GM-NN, respectively. Compared to the best message-passing architecture, PaiNN, the MAE at 110,426 training samples is 0.13 kcal/mol and 0.20 kcal/mol for PaiNN and iGM-NN$_{910}$, respectively. As seen from \figref{fig:qm9_accuracy}, the predictive accuracy of the iGM-NN model is equivalent to that of the PhysNet architecture and is better than the accuracy of the SchNet architecture.

\begin{figure*}
    \centering
    \includegraphics[width=16cm]{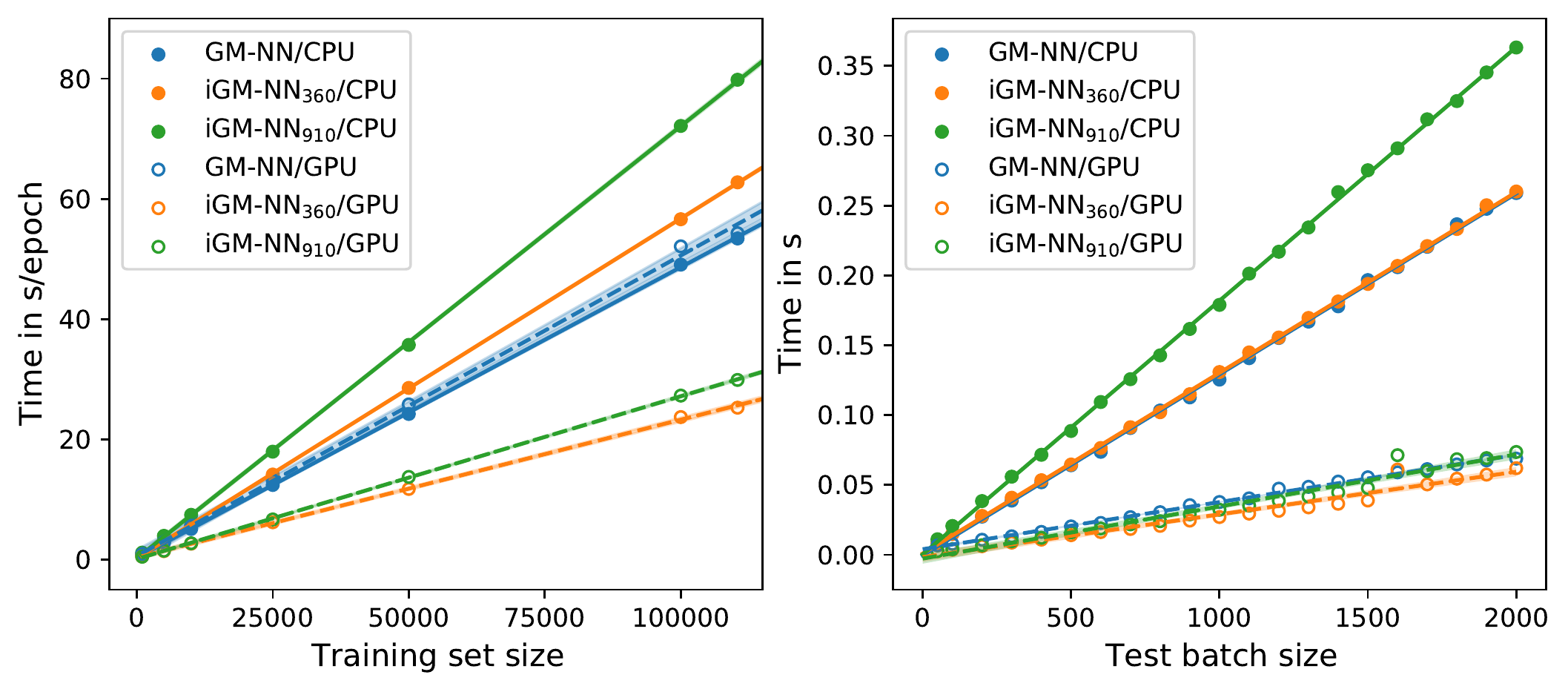}
    \caption{Training (left) and inference (right) times against the training set and test batch sizes, respectively. Linear fits are displayed for clarity and shaded areas denote the 95~\% confidence intervals for linear regression.}
    \label{fig:qm9_timing}
\end{figure*}

In the following, we compare training and inference times of the iGM-NN model with the respective values for the GM-NN model, for the PaiNN architecture, and for the kernel-based FCHL18 and FCHL19 models. The respective values for GM-NN and iGM-NN are presented in \figref{fig:qm9_timing}. The training and inference times for the respective literature methods are taken from Refs.~\citenum{Faber2018, Christensen2020, Schuett2021}.

Compared to the GM-NN model, iGM-NN$_{360}$ and iGM-NN$_{910}$ need factors of 1.8--2.2 less time for a single training epoch at a single NVIDIA Tesla V100-SXM-32GB GPU. This speedup can be explained by the parallel data loading implemented in the iGM-NN model. On a 20-core node equipped with two Intel Xeon CPU E5-2640 v4 @ 2.40GHz CPUs we observed a slightly degraded efficiency of the model. This can be explained by the inclusion of batch normalization and an overall larger model compared to GM-NN. However, the improved convergence of the iGM-NN model results in a significantly reduced number of training epochs (500 instead of 5000) and therefore leads to a much shorter overall training time compared to the previous GM-NN model.

With the training times ranging from 4.0 h (GPU) to 11 h (CPU), obtained for 110,426 samples, iGM-NN is about 3--8 times faster than the kernel-based FCHL19 model. The latter needs 27 h to calculate the kernel matrix for all structures in the QM9 data set on a 24-core node equipped with two Intel Xeon E5-2680v3 @ 2.50 GHz CPUs~\cite{Christensen2020}. Note that FCHL19 is about 20 times faster than FCHL18~\cite{Christensen2020}. The reported times to train a PhysNet model are in the range of 1--2 days~\cite{Unke2019}.

The inference times of iGM-NN and GM-NN are similar, although for the iGM-NN model a larger neural network was employed. Evaluating the iGM-NN model on a random mini-batch of 2000 structures we obtained inference time in the range of 62--74~ms, while for the GM-NN model we acquired 69~ms. Additionally, for a random mini-batch of 50 molecules from QM9, the inference time of iGM-NN models is 2.5--2.6~ms, while for the PaiNN model 13~ms were reported~\cite{Schuett2021}.

In total, we observed that the iGM-NN model performs on par with the best kernel-based methods, which have been deemed to be more data-efficient than neural networks, and with the message-passing architectures. Moreover, we find that the proposed method is about 5 times faster per inference step than PaiNN~\cite{Schuett2021}.

\subsection{\label{sec:results_md17} Results for MD17}

\begin{figure*}
\centering
\includegraphics[width=16cm]{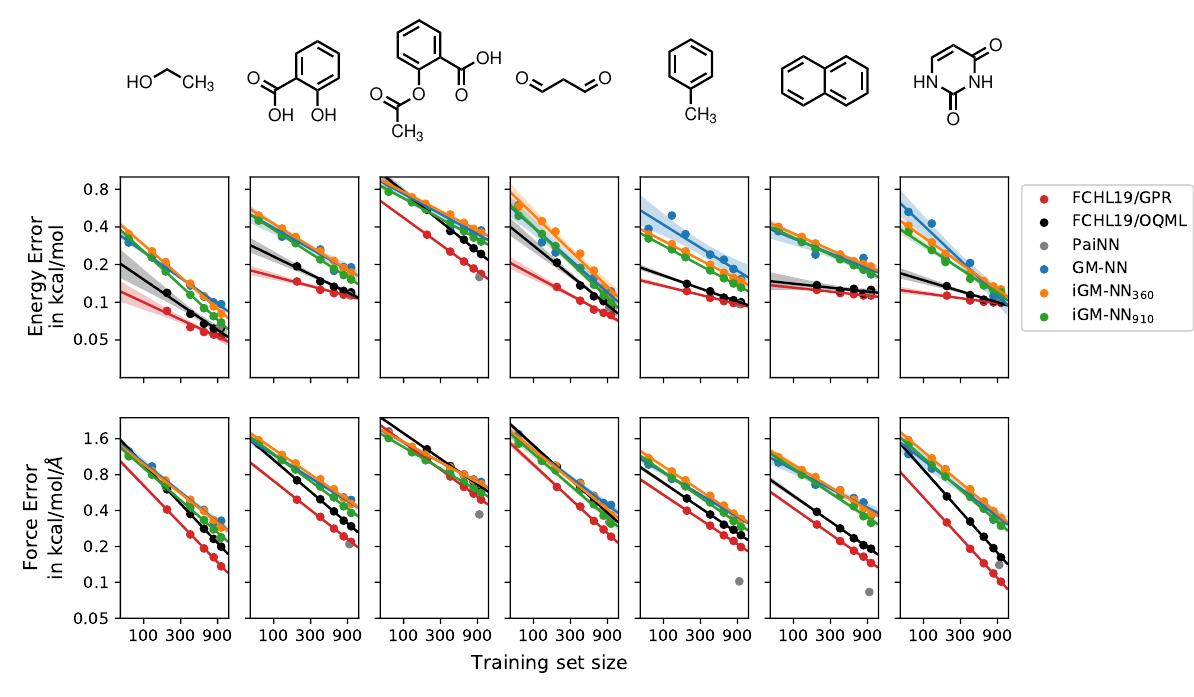}
\caption{Learning curves for seven molecules from the MD17 data set. The mean absolute errors (MAEs) of total energies and atomic forces are plotted against the training set size. Linear fits are displayed for clarity and shaded areas denote the 95~\% confidence intervals for linear regression. The respective values are given from left to right for: ethanol, salicylic acid, aspirin, malonaldehyde, toluene, naphthalene, and uracil.}
\label{fig:md17_accuracy}
\end{figure*}

We evaluate the ability to predict combined energies and forces on the MD17 benchmark. \figref{fig:md17_accuracy} reports the MAE of force and energy predictions as a function of the number of training samples taken for 7 molecules from the data set. Note that we excluded benzene from the experiments due to noisy energies. To demonstrate the data efficiency of iGM-NN, we use the more challenging training set sizes with up to 1000 structures. We compare the iGM-NN model to the kernel-based FCHL19/GPR and FCHL19/OQML approaches~\cite{Christensen2020}, which are state-of-the-art kernel-based methods. The first approach incorporates derivatives in the training set within the framework of Gaussian process regression. The second one, operator quantum machine learning (OQML), allows for simultaneous training on the energies and forces. In addition, we compare to the recently proposed equivariant message-passing architecture, PaiNN~\cite{Schuett2021}.

In general, we note that the improved GM-NN (iGM-NN) architecture leads to models that have similar or improved accuracy compared to the prediction errors reported for the GM-NN architecture in our previous paper~\cite{Zaverkin2020}. For all molecules in the MD17 data set, the iGM-NN models learn somewhat faster compared to GM-NN for both energy and force training. As a general trend, the iGM-NN$_{910}$ model needs about 20--40~\% fewer data to get the same accuracy as the GM-NN model. The iGM-NN$_{360}$ model requires the same amount or about 10--20~\% fewer data to reach the same predictive accuracy as the GM-NN model, while using fewer invariant descriptors. For example, for aspirin, an MAE force error of about 0.7 kcal/mol/\AA{} is obtained at roughly 600, 800, and 1000 samples for iGM-NN$_{910}$, iGM-NN$_{360}$, and GM-NN, respectively. Learning curves for all these models are presented in \figref{fig:md17_accuracy}.

As seen in \figref{fig:md17_accuracy}, iGM-NN provides a predictive accuracy similar to the kernel-based methods, FCHL19/GPR and FCHL19/OQML. This implies, similar to the previous section, that the proposed architecture has a sample efficiency comparable to kernel-based approaches. Note that our approach is outperformed only by the FCHL19/GPR approach, which has an unfavorable trade-off for the time-to-train compared with the iGM-NN and FCHL19/OQML approaches. The equivariant message passing architecture PaiNN trained on 950 samples shows predictive accuracy on par with or better than the FCHL19/GPR approach, however, it is less efficient at inference compared to the iGM-NN approach, see below.

\begin{figure*}
\centering
\includegraphics[width=16cm]{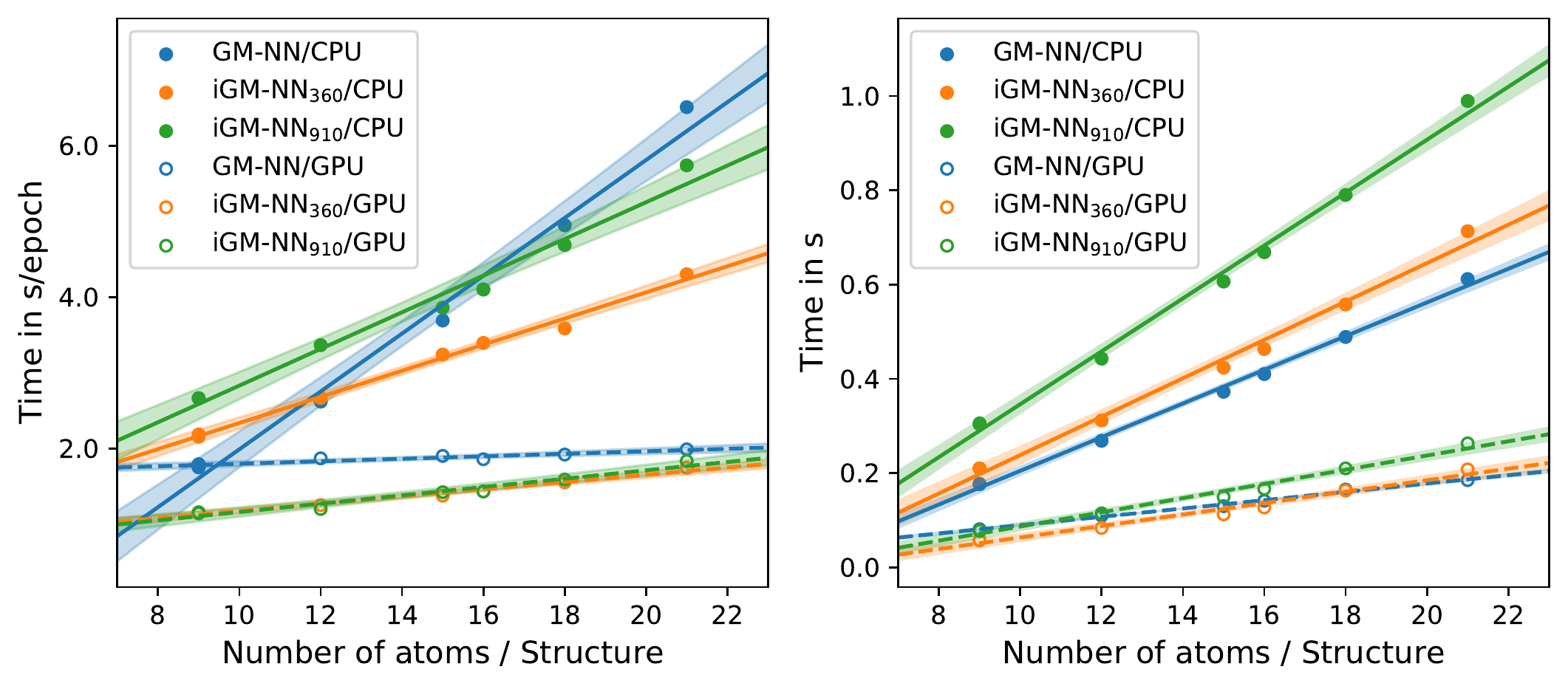}
\caption{Training (left) and inference (right) times against the training set and test batch sizes, respectively. Linear fits are displayed for clarity and shaded areas denote the 95~\% confidence intervals for linear regression. All values are given for a training set size of 1000 structures and test batch size of 2000 structures.}
\label{fig:md17_timing}
\end{figure*}

In the following, we report timings for the training and validation of the iGM-NN models for a training set of 1000 molecules and a validation batch of 2000 molecules taken from the MD17 data set. The respective values are presented in \figref{fig:md17_timing}. Compared to the GM-NN models, the speedup for a single training epoch is about a factor of 2--3 on an NVIDIA Tesla V100-SXM-32GB GPU and a 20-core node equipped with two Intel Xeon CPU E5-2640 v4 @ 2.40GHz CPUs. Additionally, a 10-fold speedup is observed due to the reduced number of training epochs (from 10,000 to 1000).

For the FCHL19/OQML model, training times vary between 51~s (malonaldehyde) and 527~s (aspirin)~\cite{Christensen2020}. The iGM-NN$_{360}$ model, with 573~s for malonaldehyde and 873~s for aspirin on a GPU, is clearly slower for smaller molecules compared to the FCHL19/OQML but comparably fast on larger ones. Additionally, the iGM-NN timing scales more favorably with the system size and the number of training samples. Compared to the FCHL19/GPR model (training time ranges from 1926~s to 101,451~s), the training of iGM-NN$_{360}$ even on a 20-core CPU node is about 2--47 times faster. The timings for kernel evaluation in the case of the FCHL19/OQML and FCHL19/GPR models were performed on a 24-core node equipped with two Intel Xeon CPU E5-2680 v3 @ 2.50GHz CPUs~\cite{Christensen2020}.


The inference time of the iGM-NN$_{360}$ model is 4.7~ms for a single ethanol and aspirin structure on a single GPU. The respective values correspond to a single energy and force evaluation times ranging from 0.5~ms/atom for ethanol to 0.2~ms/atom for aspirin, while PaiNN architecture required for a single reference calculation 14~ms (1.5~ms/atom) for ethanol and 15~ms (0.6~ms/atom) for aspirin. The inference times of the iGM-NN$_{360}$ model, evaluated on a single GPU for a mini-batch of 2000 structures, range from 57~ms for the smallest molecules, ethanol and malonaldehyde, to 208~ms for aspirin and are similar to those obtained for the GM-NN model, see \figref{fig:md17_timing}. These values correspond to a single energy and force evaluation times of 3--5~$\mu$s/atom, indicating a favorable scaling of inference times with the system size.

On a 20-core CPU node, we obtained, evaluating the  iGM-NN$_{360}$ model on single ethanol and aspirin structure, inference times ranging from 0.3~ms/atom to 0.1~ms/atom, respectively. Evaluating the respective model on a mini-batch of 2000 structures, we acquired inference times ranging from 12~$\mu$s/atom for ethanol to 17~$\mu$s/atom for aspirin. The respective force prediction times for the FCHL19/OQML model are in the range of 5.7--23.5~ms/atom~\cite{Christensen2020}.


Overall, we see that the iGM-NN, as well as GM-NN models, are at least one order of magnitude more efficient at inference compared to other state-of-the-art models. Therefore, the proposed models are favorable for atomistic simulations which require large system sizes and long time scales, e.g. molecular dynamics, while still showing comparable predictive accuracy.

\subsection{\label{sec:results_tio2} Results for Bulk \ce{TiO2}}

To determine the performance of the iGM-NN approach for a material system we investigated titanium dioxide (\ce{TiO2}). The application of \ce{TiO2} in the industry is versatile and ranges from the use as a pigment in paints~\cite{Buxbaum2008} to the use as a photocatalyst for energy-related applications, e.g. photocatalytic water splitting~\cite{Kavan1996, Khan2002}. \ce{TiO2} occurs naturally in three polymorphs: rutile, anatase, and brookite. These phases build the training set used in this section, see \secref{sec:data_tio2}.  We should note at this point that density functional theory (DFT) without semi-empirical corrections~\cite{Conesa2010, Arroyo2011}, as used to generate our training data~\cite{Artrith2016}, incorrectly predicts anatase to be more stable than rutile at standard conditions~\cite{Muscat2002, Labat2007}. Machine learning approaches cannot improve on that. Finally, to test the transferability of the iGM-NN approach, we predict properties of two high-pressure \ce{TiO2} phases, columbite and baddeleyite, which were not part of the training data.

\begin{figure*}[t]
\centering
\includegraphics[width=16cm]{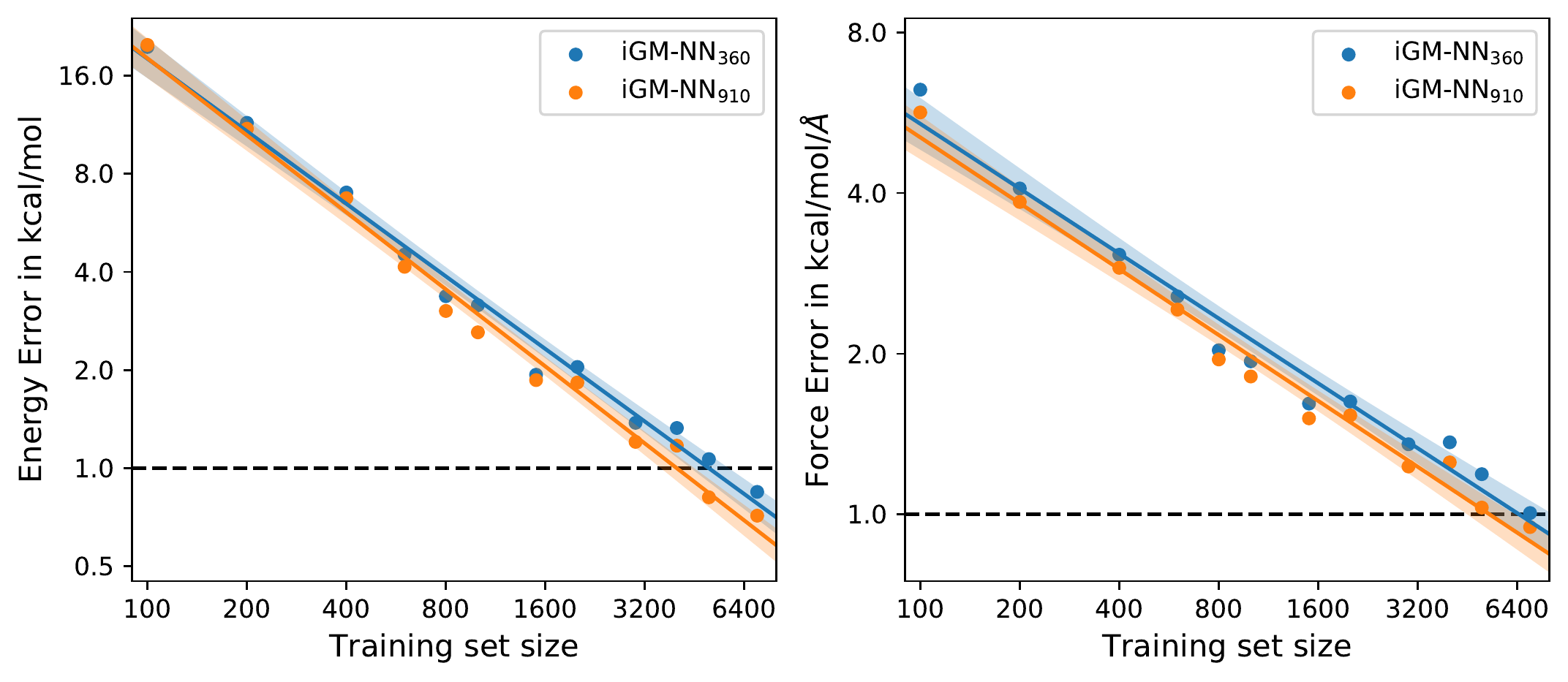}
\caption{Learning curves for the TiO$_2$ data set. The mean absolute errors (MAEs) of total energies and atomic forces are plotted against the training set size. Linear fits are displayed for clarity and shaded areas denote the 95~\% confidence intervals for linear regression. The dashed black lines represent the desired accuracies of 1 kcal/mol and 1 kcal/mol/\AA{}, respectively.}
\label{fig:fig_tio2_accuracy}
\end{figure*}

\Figref{fig:fig_tio2_accuracy} shows learning curves for the iGM-NN predictions of the total energy and atomic forces based on $360$ and $910$ local invariant molecular descriptors. As seen from the figure, the desired accuracy of 1 kcal/mol for total energies is reached for the training set size of around 4000 and 5000 reference structures, using 910 and 360 invariant features, respectively. The desired accuracy of 1 kcal/mol/\AA{} for atomic forces is reached after training on 5000 and 7000 reference structures. In general, we see that iGM-NN$_{910}$ models show only slightly lower MAE values compared to iGM-NN$_{360}$ at the cost of higher training and inference times (see previous sections). Therefore, for the following experiments, we use only machine-learned potentials (MLPs) based on 360 invariant atomic features.

\begin{figure}[t]
\centering
\includegraphics[width=8cm]{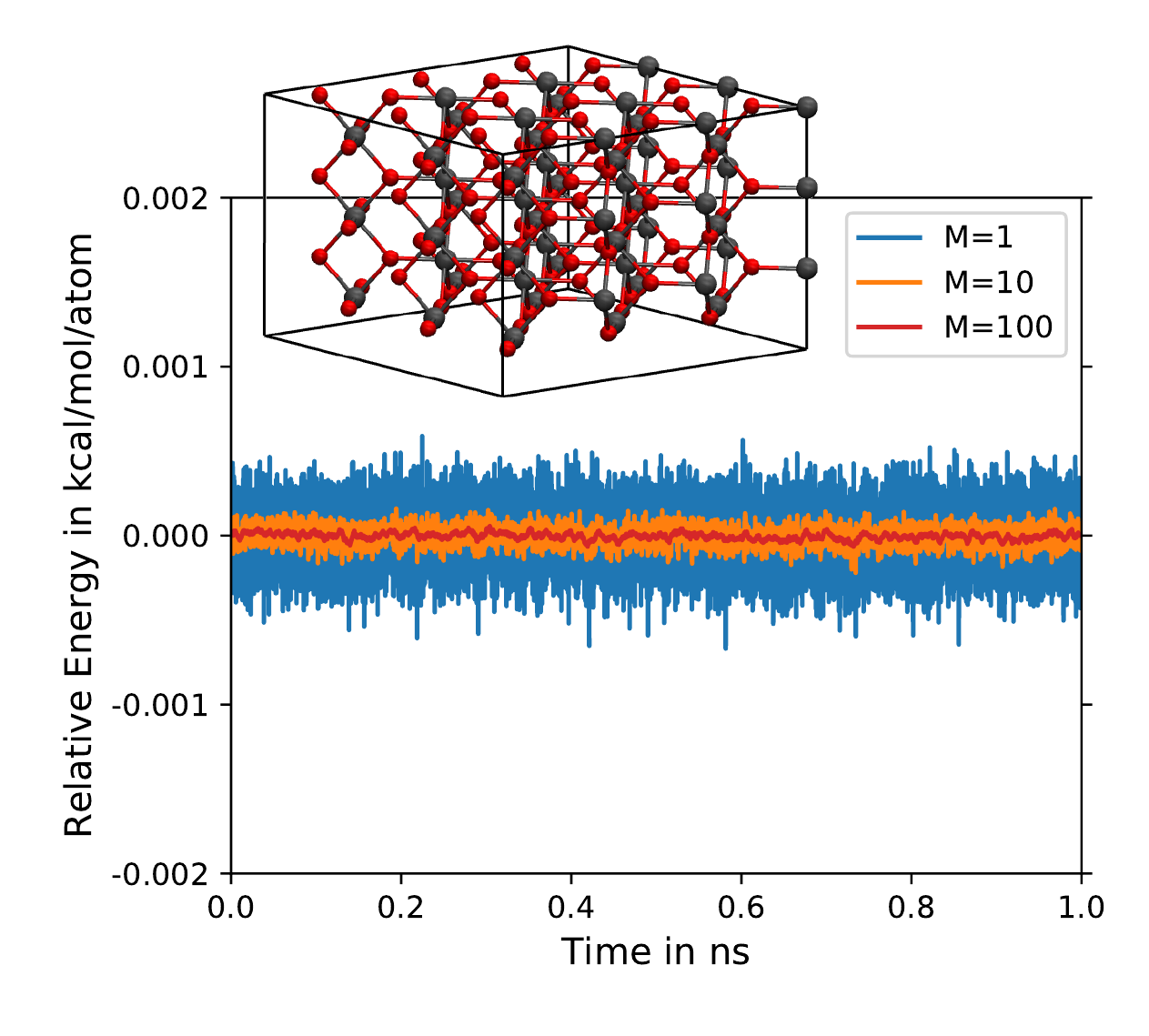}
\caption{Fluctuation of the total energy during MD simulations of a $3 \times3 \times 3$ rutile \ce{TiO2} supercell containing 162 atoms over 1~ns using a time step of 1~fs in the microcanonical (NVE) ensemble employing the iGM-NN$_{360}$ interatomic potential trained on 5000 reference structures. The atomic velocities were initialized with a Maxwell-Boltzmann distribution for a temperature of 1000~K. $M$ is the window size for the running average in \eqref{eq:running_avg}.}
\label{fig:fig_tio2_nve}
\end{figure}

The MAE of total energies and atomic forces, discussed above, are abstract quality measures of MLPs. In practice, the robustness and reliability of the potential in real-time applications provide a more rigorous assessment. Therefore, we assess the robustness and smoothness of the iGM-NN potential by applying it to a $3\times3\times3$ rutile \ce{TiO2} supercell containing 54 \ce{TiO2} formula units (162 atoms) to run a molecular dynamics (MD) simulation. Note that for an MD simulation a smooth, continuous energy surface is required to facilitate the numerical integration of the equation of motion. \Figref{fig:fig_tio2_nve} shows the total energy during MD simulation in the microcanonical (NVE) statistical ensemble, carried out within ASE simulation package~\cite{Hjorth2017} over 1.0~ns using a time step of 1.0~fs. The atomic velocities were initialized with a Maxwell-Boltzmann distribution for a temperature of 1000~K.

As seen in \Figref{fig:fig_tio2_nve}, the total energy of the system is well conserved with fluctuations below $0.5\e{-3}$ kcal/mol/atom. Since the raw data of the total energy fluctuates relatively strongly (blue line in \figref{fig:fig_tio2_nve}), we computed the running averages of the total energy to see whether the observable displays a time-drift. The running average for an observable $O_i$ can be computed as 
\begin{equation}
    \label{eq:running_avg}
    \hat{O}_i = \frac{1}{M+1} \sum_{j=i - M/2}^{i + M/2} O_j~,
\end{equation}
where $M$ is the window size of the running average. From \figref{fig:fig_tio2_nve} it can be observed that the total energy shows only a slight energy drift of $(-2.0 \pm 0.4)\e{-5}$~kcal/mol/atom/ns, which is another indication for the numerical stability of the machine-learned potential.

\begin{figure}[t]
\centering
\includegraphics[width=8cm]{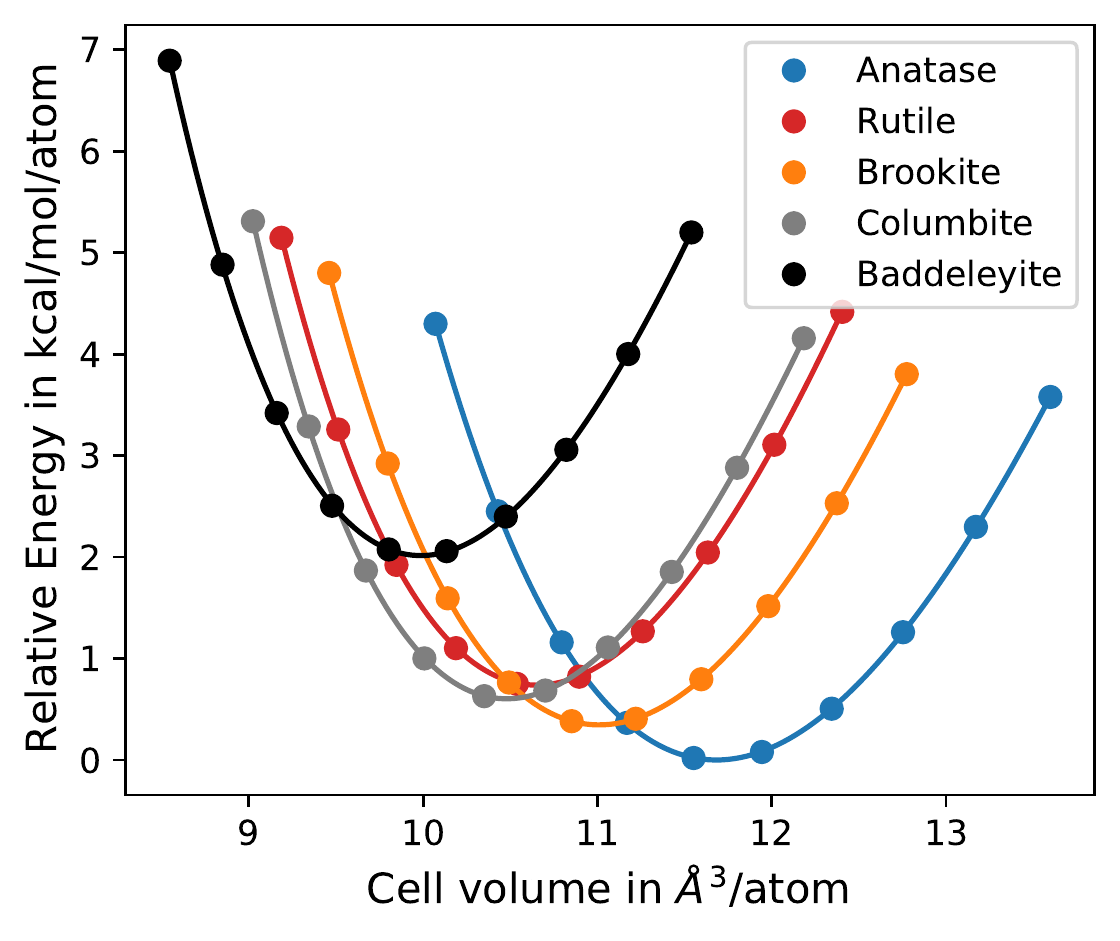}
\caption{Relative energy per atom as a function of the cell volume for the five \ce{TiO2} crystal structures: anatase, rutile, brookite, columbite, and baddeleyite.  Symbols correspond to the iGM-NN$_{360}$ potential energies and lines represent the fit of the stabilized jellium equation of state (SJEOS)~\cite{Alchagirov2003}. All energies are obtained employing the iGM-NN$_{360}$ model trained on 5000 reference structures.}
\label{fig:fig_tio2_eos}
\end{figure}

Along with the aforementioned energy conservation, we investigate whether the iGM-NN models correctly reproduce the relative phase stability. As can be seen from \figref{fig:fig_tio2_eos}, the iGM-NN potential smoothly reproduces the energy as a function of the lattice volume of rutile, anatase, and brookite. Moreover, we observe a smooth energy dependence on the lattice volume for the columbite and baddeleyite high-pressure phases which were not included in the training set. While the former exhibits similarities to anatase and rutile polymorphs, the latter is denser and each titanium is sevenfold coordinated by oxygen, as opposed to the octahedral coordination in other phases. Therefore, we clearly see that the iGM-NN approach results in smooth potentials even when extrapolating to configurations and crystal structures that have not been seen before.

To study the relative phase stability more rigorously, we computed the relative phase energies, unit cell volumes, and bulk moduli by a fit of the stabilized jellium equation of state (SJEOS)~\cite{Alchagirov2003}, as implemented in ASE, to the iGM-NN potential energy curves. The reference DFT values are taken from Ref.~\citenum{Artrith2016}, where they were computed by a fit of the Birch-Murnaghan equation of state~\cite{Birch1947} to the DFT potential energy curves. \Tabref{tab:table_tio2_eos} shows the respective values for the relative phase energies, unit cell volumes, and bulk moduli obtained using DFT and MLPs. 

\begin{table}
\caption{\label{tab:table_tio2_eos}Relative energies ($E_0$) in kcal/mol/atom, unit cell volumes ($V_0$) in \AA$^3$/atom, and bulk moduli ($B_0$) in GPa of five different \ce{TiO2} phases obtained employing the machine learned interatomic potential (iGM-NN$_{360}$) and respective reference values as computed using density-functional theory (DFT). }
\centering
\begin{tabular}{lcccccc}
\hline
\multirow{2}{*}{Phase} & \multicolumn{3}{c}{DFT} & \multicolumn{3}{c}{iGM-NN$_{360}$} \\
\cline{2-4} \cline{5-7}
    & $E_0$ & $V_0$ & $B_0$ & $E_0$ & $V_0$ & $B_0$ \\
\hline
Anatase (I$4_1$/amd) & 0.00 & 11.71 & 211 & 0.00 & 11.68 & 208\\
Rutile (P$4_2$/mnm) & 0.76 & 10.68 & 235 & 0.74 & 10.65 & 234\\
Brookite (Pbca) & 0.37 & 11.01 & 225 & 0.35 & 11.01 & 220\\
Columbite (Pbcn) & 0.67 & 10.50 & 230 & 0.60 & 10.48 & 241\\
Baddeleyite (P$2_1$/c) & 1.52 & 9.96 & 240 & 2.01 & 9.99 & 243\\
\hline
\end{tabular}
\end{table}

For three \ce{TiO2} phases that were included in the training set (rutile, anatase, and brookite), we found an excellent agreement with the reference DFT values. The relative phase energies, unit cell volumes, and bulk moduli deviate by at most 0.02~kcal/mol/atom (corresponds to max. 5.4~\%), 0.03~\AA{}$^3$/atom (0.3~\%), and 5~GPa (2.2~\%). For the high-pressure \ce{TiO2} phases columbite and baddeleyite, which are not included in the training set, we have found the relative phase energies to deviate by 0.07~kcal/mol/atom and 0.49~kcal/mol/atom. In Ref.~\citenum{Artrith2016} somewhat larger deviations of about 0.83~kcal/mol/atom and 1.20~kcal/mol/atom were observed for columbite and baddeleyite, respectively. The unit cell volumes and bulk moduli predicted by the iGM-NN potential are in excellent agreement with DFT and deviate from the reference values only by at most 0.3~\% and 4.8~\%, respectively.

In total, the obtained results demonstrate the applicability of the iGM-NN potentials to material systems. While being accurate and sample-efficient, the iGM-NN approach shows an improved transferability to polymorphs not seen during training, compared to other machine learning approaches described in the literature. This makes it feasible to study, for example in this specific case, other \ce{TiO2} phases without directly including them into the training set, and thus facilitating the study and discovery of new materials. 

\subsection{\label{sec:results_lmnto} Results for Quaternary Metal Oxide}

To assess the applicability of machine-learned potentials (MLP) based on the iGM-NN approach to a material with a more complex chemical composition we finally studied Li-Mo-Ni-Ti oxide (LMNTO, \ce{Li8Mo2Ni7Ti7O32}). The respective data set is described in \secref{sec:data_lmnto} and in Ref.~\citenum{Cooper2020}. Note that LMNTO is a potential high-capacity positive electrode material for lithium-ion batteries and, therefore, is of high technological relevance~\cite{Lee2015}.

\begin{figure*}
\centering
\includegraphics[width=16cm]{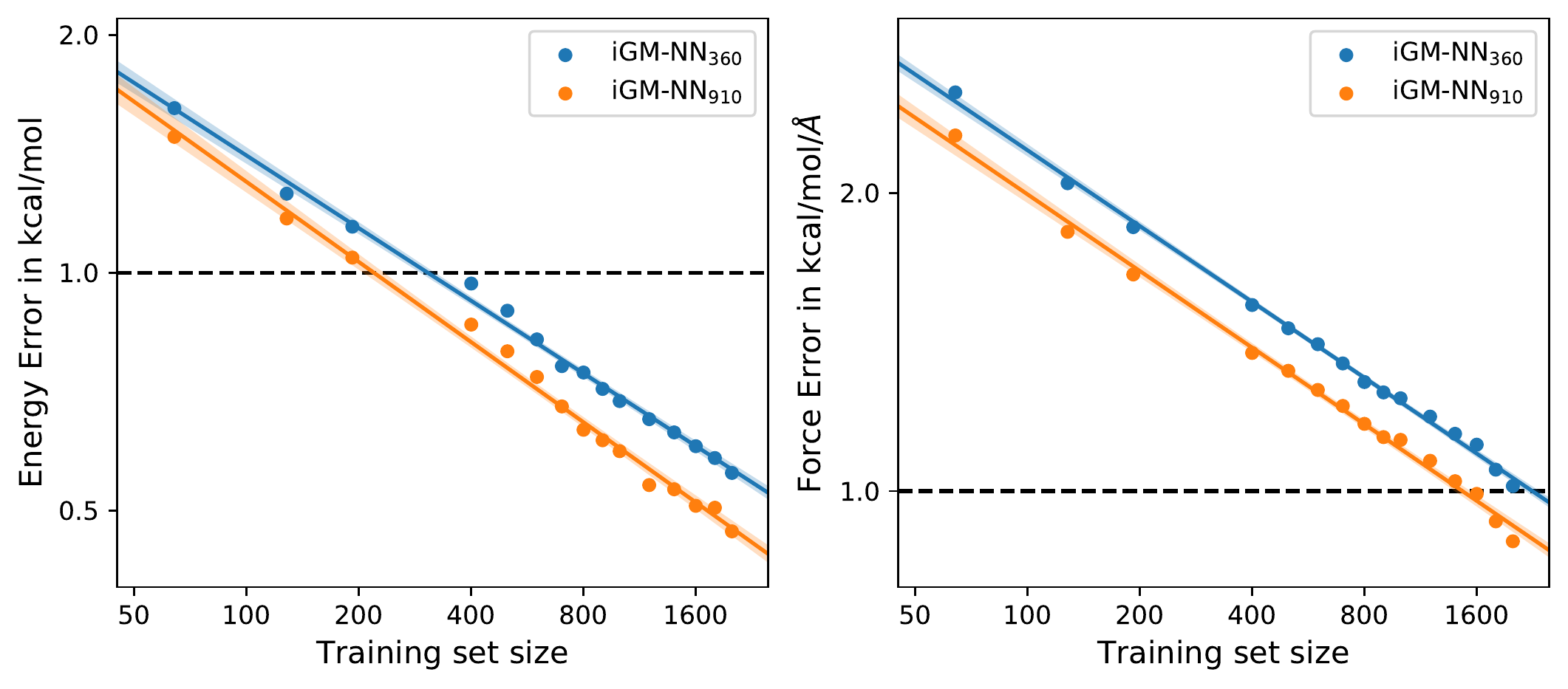}
\caption{Learning curves for the LMNTO data set. The mean absolute errors (MAEs) of total energies (left) and atomic forces (right) are plotted against the training set size. Linear fits are displayed for clarity and shaded areas denote the 95~\% confidence intervals for linear regression. The dashed black lines represent the desired accuracies of 1 kcal/mol and 1 kcal/mol/\AA{}, respectively.}
\label{fig:fig_lmnto_accuracy}
\end{figure*}

\Figref{fig:fig_lmnto_accuracy} depicts learning curves of  the total energy and atomic forces for the iGM-NN$_{360}$ and iGM-NN$_{910}$ models. 
For this system, we can compare our iGM-NN approach directly to the \ae{}net code.
The MAE for total energies at 700 training samples is 0.01~kcal/mol/atom, 0.01~kcal/mol/atom, and 0.11~kcal/mol/atom for iGM-NN$_{360}$, iGM-NN$_{910}$, and \ae{}net~\cite{Artrith2016, Artrith2017, Cooper2020}, respectively. For the MAE of atomic forces we obtained a value of 1.35~kcal/mol/\AA{} and 1.22~kcal/mol/\AA{} with MLP based on iGM-NN$_{360}$ and iGM-NN$_{910}$, respectively, trained on 700 reference structures, while \ae{}net results in an MAE of 15.22~kcal/mol/\AA{}. Note, that the \ae{}net potential was trained using the recently proposed Taylor-expansion data enhancement approach~\cite{Cooper2020}, which is more data-efficient, i.e. it uses less than 50~\% of the information contained in the data set.

\begin{figure}
\centering
\includegraphics[width=8cm]{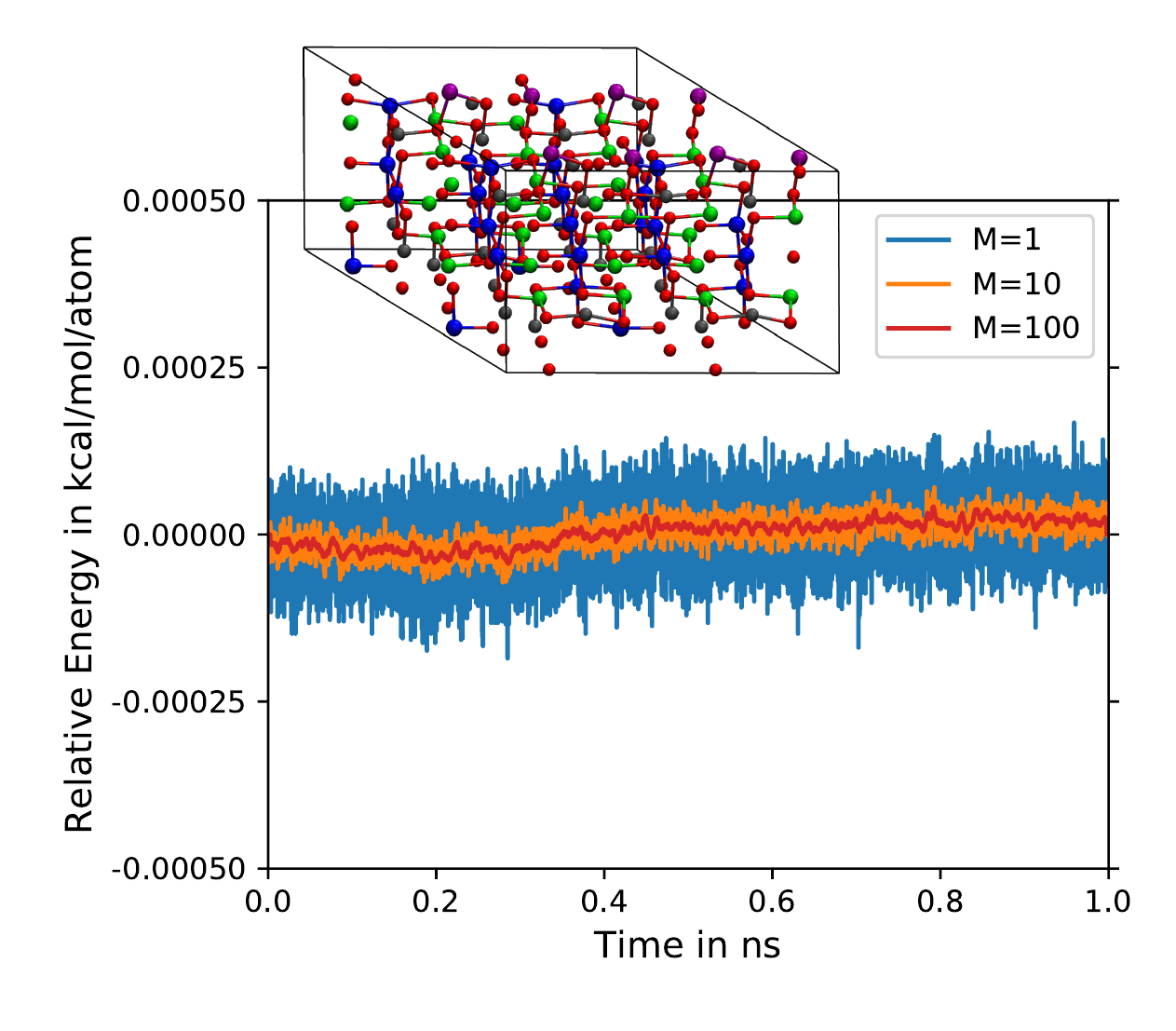}
\caption{Fluctuation of the total energy during MD simulations of a $2 \times2 \times 1$ LMNTO supercell containing 224 atoms over 1~ns using a time step of 1~fs in the microcanonical (NVE) ensemble employing the iGM-NN$_{360}$ interatomic potential trained on 2000 reference structures. The atomic velocities were initialized with a Maxwell-Boltzmann distribution for a temperature of 300~K. $M$ is the window size for the running average in \eqref{eq:running_avg}.}
\label{fig:fig_lmnto_energy_conservation}
\end{figure}

As can be seen from \figref{fig:fig_lmnto_accuracy}, both iGM-NN$_{360}$ and iGM-NN$_{910}$ reach the desired accuracy in predicted total energies already after training on 400 and 200 reference structures, respectively. The MAE target in predicted atomic forces of 1~kcal/mol/\AA{} is reached after training on 2000 and 1400 reference structures for iGM-NN$_{360}$ and iGM-NN$_{910}$, respectively. However, because of the similar performance of iGM-NN$_{360}$ and iGM-NN$_{910}$, in the following experiments we use only iGM-NN$_{360}$, similar to \secref{sec:results_tio2}. 

As discussed in \secref{sec:results_tio2}, the robustness and reliability of the MLP in a real-time simulation is a better assessment of its quality than abstract measures such as MAE values for predictions. \Figref{fig:fig_lmnto_energy_conservation} shows the total energy during an MD simulation in the microcanonical (NVE) statistical ensemble, carried out within the ASE simulation package~\cite{Hjorth2017} over 1.0~ns using a time step of 1.0~fs. The total energy of the system is well-conserved and fluctuations remain below $0.1\e{-3}$ kcal/mol/atom. Another indication for the numerical stability of the MLP based on the iGM-NN approach is very low energy drift of merely $(5.9 \pm 0.1)\e{-5}$~kcal/mol/atom/ns.

Overall, from the above experiments, we have seen that the iGM-NN approach is able to produce reliable and robust machine-learned potentials for multi-component periodic systems. Additionally, we can argue that the proposed approach shows an excellent transferability since it allowed to run a stable MD simulation for 1~ns after training on only 2000 reference structures. Note, that usually it is stated that training sets with tens of thousands of reference structures are required to construct accurate MLPs~\cite{Cooper2020}.

\section{\label{sec:conclusions} Conclusions}

We have presented an improved version of the Gaussian moment neural network~\cite{Zaverkin2020} which enables the generation of machine-learned potentials with an accuracy comparable to the established machine learning models. The presented modifications to the neural network architecture, the training process, and the implementation allow for much faster model training compared to the previous version in Ref.~\citenum{Zaverkin2020} and extends the applicability of Gaussian Moments to periodic structures. Fast training is a pre-requisite of workflows with frequent retraining, such as active learning or learning-on-the-fly, which in turn reduce the number of required ab-initio calculations.

Machine-learned potentials evaluated on two standard benchmark data sets, QM9 and MD17, have shown prediction accuracy comparable to state-of-the-art machine learning models. For models trained on atomization energies only, such as QM9, the accuracy is close to 0.2~kcal/mol, and thus approaches the accuracy of message-passing models, such as PhysNet. The training time was reduced by factors of 10--20  compared to GM-NN.

Testing the proposed approach on the MD17 data set, which contains total energies and atomic forces, we have observed only a slight improvement in the prediction accuracy but a considerable reduction (factors of 10--20) of training times compared to the previous GM-NN. Note that GM-NN already could achieve an accuracy comparable to the state-of-the-art machine learning models. We should emphasize that the iGM-NN models outperform current state-of-the-art literature ML methods at inference time in terms of time required for single energy and force calculation.

We have shown that the iGM-NN approach can be used to generate accurate and robust machine-learned potentials for periodic systems, such as titanium dioxide (\ce{TiO2}) and LMNTO (\ce{Li8Mo2Ni7Ti7O32}). For both systems, the proposed approach could achieve the desired accuracy of 1~kcal/mol and 1~kcal/mol/\AA{} in predicted total energies and atomic forces, respectively. Additionally, we assessed the quality of the machine-learned potentials by inspecting the energy conservation during a microcanonical molecular dynamics simulation. Along with energy fluctuations of the order of $10^{-4}$~kcal/mol/atom we observed negligible time-drift, which demonstrates the smoothness and robustness of the iGM-NN potentials.

Testing the relative stability of \ce{TiO2} polymorphs, we have observed an excellent transferability of the iGM-NN potentials. Specifically, we could obtain an accurate prediction for relative energies, unit cell volumes, and bulk moduli on high-pressure \ce{TiO2} phases columbite and baddeleyite, which were not given in the training set.

In summary, the developments of this work aim to improve the applicability of machine-learned potentials to run various simulations for molecules and materials. Future work will also deal with long-range interactions important for, e.g., ionic liquids.

\section*{Acknowledgement}

We thank the Deutsche Forschungsgemeinschaft (DFG, German Research Foundation) for supporting this work by funding EXC 2075 - 390740016 under Germany's Excellence Strategy. We acknowledge the support by the Stuttgart Center for Simulation Science (SimTech) and the European Union's Horizon 2020 research and innovation programme (grant agreement No. 646717, TUNNELCHEM). The authors acknowledge support by the state of Baden-Württemberg through the bwHPC consortium for providing computer and GPU time. The authors thank the International Max Planck Research School for Intelligent Systems (IMPRS-IS)
for supporting David Holzmüller. Viktor Zaverkin acknowledges the financial support received in the form of a PhD scholarship from the Studienstiftung des  Deutschen  Volkes (German National Academic Foundation).

\bibliography{main}

\begin{thebibliography}{10}

\bibitem{Hornak06}
V.~Hornak, R.~Abel, A.~Okur, B.~Strockbine, A.~Roitberg, and C.~Simmerling,
  ``Comparison of multiple amber force fields and development of improved
  protein backbone parameters,'' {\em Proteins}, vol.~65, no.~3, pp.~712--725,
  2006.

\bibitem{Vanommeslaeghe10}
K.~Vanommeslaeghe, E.~Hatcher, C.~Acharya, S.~Kundu, S.~Zhong, J.~Shim,
  E.~Darian, O.~Guvench, P.~Lopes, I.~Vorobyov, and A.~D. Mackerell~Jr.,
  ``Charmm general force field: A force field for drug-like molecules
  compatible with the charmm all-atom additive biological force fields,'' {\em
  J. Comput. Chem}, vol.~31, no.~4, pp.~671--690, 2010.

\bibitem{Halgren96}
T.~A. Halgren, ``Merck molecular force field. i. basis, form, scope,
  parameterization, and performance of mmff94,'' {\em J. Comput. Chem.},
  vol.~17, no.~5‐6, pp.~490--519, 1996.

\bibitem{Behler2010}
J.~Behler, ``Atom-centered symmetry functions for constructing high-dimensional
  neural network potentials,'' {\em J. Chem. Phys.}, vol.~134, no.~7,
  p.~074106, 2011.

\bibitem{Bartok2010}
A.~P. Bart\'ok, M.~C. Payne, R.~Kondor, and G.~Cs\'anyi, ``Gaussian
  approximation potentials: The accuracy of quantum mechanics, without the
  electrons,'' {\em Phys. Rev. Lett.}, vol.~104, p.~136403, 2010.

\bibitem{Rupp2012}
M.~Rupp, A.~Tkatchenko, K.-R. M\"uller, and O.~A. von Lilienfeld, ``Fast and
  accurate modeling of molecular atomization energies with machine learning,''
  {\em Phys. Rev. Lett.}, vol.~108, p.~058301, 2012.

\bibitem{Bartok2013}
A.~P. Bart\'ok, R.~Kondor, and G.~Cs\'anyi, ``On representing chemical
  environments,'' {\em Phys. Rev. B}, vol.~87, p.~184115, 2013.

\bibitem{Linienfeld2015}
O.~A. von Lilienfeld, R.~Ramakrishnan, M.~Rupp, and A.~Knoll, ``Fourier series
  of atomic radial distribution functions: A molecular fingerprint for machine
  learning models of quantum chemical properties,'' {\em Int. J. Quantum
  Chem.}, vol.~115, no.~16, pp.~1084--1093, 2015.

\bibitem{Shapeev2016}
A.~V. Shapeev, ``Moment tensor potentials: A class of systematically improvable
  interatomic potentials,'' {\em Multiscale Model. Simul.}, vol.~14, no.~3,
  pp.~1153--1173, 2016.

\bibitem{Khorshidi2016}
A.~Khorshidi and A.~A. Peterson, ``Amp: A modular approach to machine learning
  in atomistic simulations,'' {\em Comput. Phys. Commun.}, vol.~207, pp.~310 --
  324, 2016.

\bibitem{Schuett2017}
K.~Sch\"{u}tt, P.-J. Kindermans, H.~E. Sauceda~Felix, S.~Chmiela,
  A.~Tkatchenko, and K.-R. M\"{u}ller, ``Schnet: A continuous-filter
  convolutional neural network for modeling quantum interactions,'' in {\em
  NeurIPS 30} (I.~Guyon, U.~V. Luxburg, S.~Bengio, H.~Wallach, R.~Fergus,
  S.~Vishwanathan, and R.~Garnett, eds.), pp.~991--1001, Curran Associates,
  Inc., 2017.

\bibitem{Artrith2017}
N.~Artrith, A.~Urban, and G.~Ceder, ``Efficient and accurate machine-learning
  interpolation of atomic energies in compositions with many species,'' {\em
  Phys. Rev. B}, vol.~96, p.~014112, 2017.

\bibitem{Faber2018}
F.~A. Faber, A.~S. Christensen, B.~Huang, and O.~A. von Lilienfeld,
  ``Alchemical and structural distribution based representation for universal
  quantum machine learning,'' {\em J. Chem. Phys.}, vol.~148, no.~24,
  p.~241717, 2018.

\bibitem{Zhang2018}
L.~Zhang, J.~Han, H.~Wang, R.~Car, and W.~E, ``Deep potential molecular
  dynamics: A scalable model with the accuracy of quantum mechanics,'' {\em
  Phys. Rev. Lett.}, vol.~120, p.~143001, Apr 2018.

\bibitem{Zhang2018_2}
L.~Zhang, J.~Han, H.~Wang, W.~Saidi, R.~Car, and W.~E, ``{End-to-end Symmetry
  Preserving Inter-atomic Potential Energy Model for Finite and Extended
  Systems},'' in {\em NeuRIPS} (S.~Bengio, H.~Wallach, H.~Larochelle,
  K.~Grauman, N.~Cesa-Bianchi, and R.~Garnett, eds.), vol.~31, Curran
  Associates, Inc., 2018.

\bibitem{Kocer2019}
E.~Kocer, J.~K. Mason, and H.~Erturk, ``A novel approach to describe chemical
  environments in high-dimensional neural network potentials,'' {\em J. Chem.
  Phys.}, vol.~150, no.~15, p.~154102, 2019.

\bibitem{Zhang2019}
Y.~Zhang, C.~Hu, and B.~Jiang, ``Embedded atom neural network potentials:
  Efficient and accurate machine learning with a physically inspired
  representation,'' {\em J. Phys. Chem. Lett.}, vol.~10, no.~17,
  pp.~4962--4967, 2019.

\bibitem{Christensen2020}
A.~S. Christensen, L.~A. Bratholm, F.~A. Faber, and O.~von Lilienfeld, ``Fchl
  revisited: Faster and more accurate quantum machine learning,'' {\em J. Chem.
  Phys.}, vol.~152, no.~4, p.~44107, 2020.

\bibitem{Zaverkin2020}
V.~Zaverkin and J.~Kästner, ``Gaussian moments as physically inspired
  molecular descriptors for accurate and scalable machine learning
  potentials,'' {\em J. Chem. Theory Comput.}, vol.~16, no.~8, pp.~5410--5421,
  2020.

\bibitem{Schuett2021}
K.~T. Schütt, O.~T. Unke, and M.~Gastegger, ``Equivariant message passing for
  the prediction of tensorial properties and molecular spectra,'' {\em ArXiv},
  vol.~abs/2102.03150, 2021.

\bibitem{Behler2007}
J.~Behler and M.~Parrinello, ``Generalized neural-network representation of
  high-dimensional potential-energy surfaces,'' {\em Phys. Rev. Lett.},
  vol.~98, p.~146401, 2007.

\bibitem{Behler2011}
J.~Behler, ``Neural network potential-energy surfaces in chemistry: a tool for
  large-scale simulations,'' {\em Phys. Chem. Chem. Phys.}, vol.~13,
  pp.~17930--17955, 2011.

\bibitem{Artrith2016}
N.~Artrith and A.~Urban, ``An implementation of artificial neural-network
  potentials for atomistic materials simulations: Performance for tio2,'' {\em
  Comput. Mater. Sci.}, vol.~114, pp.~135--150, 2016.

\bibitem{Schuett2017_2}
K.~T. Schütt, F.~Arbabzadah, S.~Chmiela, K.~R. Müller, and A.~Tkatchenko,
  ``Quantum-chemical insights from deep tensor neural networks,'' {\em Nat.
  Commun.}, vol.~8, p.~13890, 2017.

\bibitem{Chmiela2017}
S.~Chmiela, A.~Tkatchenko, H.~E. Sauceda, I.~Poltavsky, K.~T. Sch{\"u}tt, and
  K.-R. M{\"u}ller, ``Machine learning of accurate energy-conserving molecular
  force fields,'' {\em Sci. Adv.}, vol.~3, no.~5, 2017.

\bibitem{Chmiela2018}
S.~Chmiela, H.~E. Sauceda, K.-R. Müller, and A.~Tkatchenko, ``Towards exact
  molecular dynamics simulations with machine-learned force fields,'' {\em Nat.
  Commun.}, vol.~9, p.~3887, 2018.

\bibitem{Gubaev2018}
K.~Gubaev, E.~V. Podryabinkin, and A.~V. Shapeev, ``Machine learning of
  molecular properties: Locality and active learning,'' {\em J. Chem. Phys.},
  vol.~148, no.~24, p.~241727, 2018.

\bibitem{Lubbers2018}
N.~Lubbers, J.~S. Smith, and K.~Barros, ``Hierarchical modeling of molecular
  energies using a deep neural network,'' {\em J. Chem. Phys.}, vol.~148,
  no.~24, p.~241715, 2018.

\bibitem{Yao2018}
K.~Yao, J.~E. Herr, D.~W. Toth, R.~Mckintyre, and J.~Parkhill, ``The
  tensormol-0.1 model chemistry: a neural network augmented with long-range
  physics,'' {\em Chem. Sci.}, vol.~9, pp.~2261--2269, 2018.

\bibitem{Unke2019}
O.~T. Unke and M.~Meuwly, ``Physnet: A neural network for predicting energies,
  forces, dipole moments, and partial charges,'' {\em J. Chem. Theory Comput.},
  vol.~15, no.~6, pp.~3678--3693, 2019.

\bibitem{Cooper2020}
A.~M. Cooper, J.~Kästner, A.~Urban, and N.~Artrith, ``Efficient training of
  ann potentials by including atomic forces via taylor expansion and
  application to water and a transition-metal oxide,'' {\em npj Comput.
  Mater.}, vol.~6, no.~54, 2020.

\bibitem{Smith2019}
J.~S. Smith, B.~T. Nebgen, R.~Zubatyuk, N.~Lubbers, C.~Devereux, K.~Barros,
  S.~Tretiak, O.~Isayev, and A.~E. Roitberg, ``{Approaching coupled cluster
  accuracy with a general-purpose neural network potential through transfer
  learning},'' {\em Nat. Commun.}, vol.~10, no.~1, p.~2903, 2019.

\bibitem{Sivaraman2021}
G.~Sivaraman, L.~Gallington, A.~N. Krishnamoorthy, M.~Stan, G.~Cs\'anyi,
  A.~V\'azquez-Mayagoitia, and C.~J. Benmore, ``Experimentally driven automated
  machine-learned interatomic potential for a refractory oxide,'' {\em Phys.
  Rev. Lett.}, vol.~126, p.~156002, Apr 2021.

\bibitem{Mackerell2004}
A.~D. Mackerell~Jr., ``Empirical force fields for biological macromolecules:
  Overview and issues,'' {\em J. Comput. Chem.}, vol.~25, no.~13,
  pp.~1584--1604, 2004.

\bibitem{Hornik1991}
K.~Hornik, ``Approximation capabilities of multilayer feedforward networks,''
  {\em Neural Netw.}, vol.~4, no.~2, pp.~251 -- 257, 1991.

\bibitem{Blank1995}
T.~B. Blank, S.~D. Brown, A.~W. Calhoun, and D.~J. Doren, ``{Neural network
  models of potential energy surfaces},'' {\em J. Chem. Phys.}, vol.~103,
  no.~10, pp.~4129--4137, 1995.

\bibitem{Lorenz2004}
S.~Lorenz, A.~Gro{\ss}, and M.~Scheffler, ``{Representing high-dimensional
  potential-energy surfaces for reactions at surfaces by neural networks},''
  {\em Chem. Phys. Lett.}, vol.~395, no.~4, pp.~210--215, 2004.

\bibitem{Settles2009}
B.~Settles, ``Active learning literature survey,'' Computer Sciences Technical
  Report 1648, University of Wisconsin--Madison, 2009.

\bibitem{Li15}
Z.~Li, J.~R. Kermode, and A.~De~Vita, ``Molecular dynamics with on-the-fly
  machine learning of quantum-mechanical forces,'' {\em Phys. Rev. Lett.},
  vol.~114, p.~096405, 2015.

\bibitem{Podryabinkin17}
E.~V. Podryabinkin and A.~V. Shapeev, ``Active learning of linearly
  parametrized interatomic potentials,'' {\em Comput. Mater. Sci.}, vol.~140,
  pp.~171 -- 180, 2017.

\bibitem{Smith18}
J.~S. Smith, B.~Nebgen, N.~Lubbers, O.~Isayev, and A.~E. Roitberg, ``Less is
  more: Sampling chemical space with active learning,'' {\em J. Chem. Phys.},
  vol.~148, no.~24, p.~241733, 2018.

\bibitem{Zhang19}
L.~Zhang, D.-Y. Lin, H.~Wang, R.~Car, and W.~E, ``Active learning of uniformly
  accurate interatomic potentials for materials simulation,'' {\em Phys. Rev.
  Materials}, vol.~3, p.~023804, Feb 2019.

\bibitem{Gastegger17}
M.~Gastegger, J.~Behler, and P.~Marquetand, ``Machine learning molecular
  dynamics for the simulation of infrared spectra,'' {\em Chem. Sci.}, vol.~8,
  pp.~6924--6935, 2017.

\bibitem{Zaverkin2021}
V.~Zaverkin and J.~K\"astner, ``Exploration of transferable and uniformly
  accurate neural network interatomic potentials using optimal experimental
  design,'' {\em Mach. Learn.: Sci. Technol.}, vol.~2, p.~035009, 2021.

\bibitem{Janet19}
J.~P. Janet, C.~Duan, T.~Yang, A.~Nandy, and H.~J. Kulik, ``A quantitative
  uncertainty metric controls error in neural network-driven chemical
  discovery,'' {\em Chem. Sci.}, vol.~10, pp.~7913--7922, 2019.

\bibitem{Vandermause20}
J.~Vandermause, S.~B. Torrisi, S.~Batzner, Y.~Xie, L.~Sun, A.~M. Kolpak, and
  B.~Kozinsky, ``On-the-fly active learning of interpretable bayesian force
  fields for atomistic rare events,'' {\em Npj Comput. Mater.}, vol.~6, no.~20,
  2020.

\bibitem{Schuett2019}
K.~T. Schütt, P.~Kessel, M.~Gastegger, K.~A. Nicoli, A.~Tkatchenko, and K.-R.
  Müller, ``Schnetpack: A deep learning toolbox for atomistic systems,'' {\em
  J. Chem. Theory Comput.}, vol.~15, no.~1, pp.~448--455, 2019.

\bibitem{Ruddigkeit2012}
L.~Ruddigkeit, R.~van Deursen, L.~C. Blum, and J.-L. Reymond, ``Enumeration of
  166 billion organic small molecules in the chemical universe database
  gdb-17,'' {\em J. Chem. Inf. Model.}, vol.~52, no.~11, pp.~2864--2875, 2012.

\bibitem{Ramakrishnan2014}
R.~Ramakrishnan, P.~O. Dral, M.~Rupp, and O.~A. von Lilienfeld, ``Quantum
  chemistry structures and properties of 134 kilo molecules,'' {\em Sci. Data},
  vol.~1, no.~1, p.~140022, 2014.

\bibitem{Jacot2018}
A.~Jacot, F.~Gabriel, and C.~Hongler, ``{Neural Tangent Kernel: Convergence and
  Generalization in Neural Networks},'' in {\em NeurIPS} (S.~Bengio,
  H.~Wallach, H.~Larochelle, K.~Grauman, N.~Cesa-Bianchi, and R.~Garnett,
  eds.), vol.~31, Curran Associates, Inc., 2018.

\bibitem{Ioffe2015}
S.~Ioffe and C.~Szegedy, ``Batch normalization: Accelerating deep network
  training by reducing internal covariate shift,'' in {\em PMLR} (F.~Bach and
  D.~Blei, eds.), vol.~37, pp.~448--456, PMLR, 2015.

\bibitem{Lee2020}
J.~Lee, L.~Xiao, S.~S. Schoenholz, Y.~Bahri, R.~Novak, J.~Sohl-Dickstein, and
  J.~Pennington, ``{Wide neural networks of any depth evolve as linear models
  under gradient descent},'' {\em J. Stat. Mech.: Theory Exp.}, vol.~2020,
  no.~12, p.~124002, 2020.

\bibitem{Chizat2019}
L.~Chizat, E.~Oyallon, and F.~Bach, ``On lazy training in differentiable
  programming,'' {\em NeurIPS}, vol.~32, pp.~2937--2947, 2019.

\bibitem{Nonnenmacher2020}
M.~Nonnenmacher, D.~Reeb, and I.~Steinwart, ``Which minimizer does my neural
  network converge to?,'' {\em ArXiv}, vol.~abs/2011.02408, 2020.

\bibitem{He2015}
K.~He, X.~Zhang, S.~Ren, and J.~Sun, ``Delving deep into rectifiers: Surpassing
  human-level performance on imagenet classification,'' {\em 2015 IEEE ICCV},
  pp.~1026--1034, 2015.

\bibitem{Hendrycks2016}
D.~Hendrycks and K.~Gimpel, ``Gaussian error linear units (gelus),'' {\em
  arXiv}, vol.~abs/1606.08415, 2016.

\bibitem{Elfwing2018}
S.~Elfwing, E.~Uchibe, and K.~Doya, ``Sigmoid-weighted linear units for neural
  network function approximation in reinforcement learning,'' {\em Neural
  Netw.}, vol.~107, pp.~3--11, 2018.

\bibitem{Ramachandran2018}
P.~Ramachandran, B.~Zoph, and Q.~V. Le, ``Searching for activation functions,''
  {\em ArXiv}, vol.~abs/1710.05941, 2018.

\bibitem{Klambauer2017}
G.~Klambauer, T.~Unterthiner, A.~Mayr, and S.~Hochreiter, ``Self-normalizing
  neural networks,'' in {\em NeurIPS} (I.~Guyon, U.~V. Luxburg, S.~Bengio,
  H.~Wallach, R.~Fergus, S.~Vishwanathan, and R.~Garnett, eds.), vol.~30,
  Curran Associates, Inc., 2017.

\bibitem{Arora2019}
S.~Arora, S.~Du, W.~Hu, Z.~Li, R.~Salakhutdinov, and R.~Wang, ``On exact
  computation with an infinitely wide neural net,'' in {\em NeurIPS}, 2019.

\bibitem{Lu2020}
Y.~Lu, S.~Gould, and T.~Ajanthan, ``Bidirectional self-normalizing neural
  networks,'' {\em ArXiv}, vol.~abs/2006.12169, 2020.

\bibitem{Adam2015}
D.~P. Kingma and J.~Ba, ``Adam: A method for stochastic optimization,'' {\em
  CoRR}, vol.~abs/1412.6980, 2015.

\bibitem{Prechelt2012}
L.~Prechelt, {\em Early Stopping --- But When?}, pp.~53--67.
\newblock Berlin, Heidelberg: Springer Berlin Heidelberg, 2012.

\bibitem{Perdew1996}
J.~P. Perdew, K.~Burke, and M.~Ernzerhof, ``Generalized gradient approximation
  made simple,'' {\em Phys. Rev. Lett.}, vol.~77, pp.~3865--3868, 1996.

\bibitem{Giannozzi2009}
P.~Giannozzi, S.~Baroni, N.~Bonini, M.~Calandra, R.~Car, C.~Cavazzoni,
  D.~Ceresoli, G.~L. Chiarotti, M.~Cococcioni, I.~Dabo, A.~D. Corso,
  S.~de~Gironcoli, S.~Fabris, G.~Fratesi, R.~Gebauer, U.~Gerstmann,
  C.~Gougoussis, A.~Kokalj, M.~Lazzeri, L.~Martin-Samos, N.~Marzari, F.~Mauri,
  R.~Mazzarello, S.~Paolini, A.~Pasquarello, L.~Paulatto, C.~Sbraccia,
  S.~Scandolo, G.~Sclauzero, A.~P. Seitsonen, A.~Smogunov, P.~Umari, and R.~M.
  Wentzcovitch, ``{{QUANTUM} {ESPRESSO}: a modular and open-source software
  project for quantum simulations of materials},'' {\em J. Phys. Condens.
  Matter}, vol.~21, no.~39, p.~395502, 2009.

\bibitem{Sun2015}
J.~Sun, A.~Ruzsinszky, and J.~P. Perdew, ``Strongly constrained and
  appropriately normed semilocal density functional,'' {\em Phys. Rev. Lett.},
  vol.~115, p.~036402, 2015.

\bibitem{Cooper2020_2}
A.~Cooper, J.~Kästner, A.~Urban, and N.~Artrith, ``Efficient training of ann
  potentials by including atomic forces via taylor expansion and application to
  water and a transition-metal oxide,'' 2020.

\bibitem{Buxbaum2008}
G.~Buxbaum, {\em Industrial Inorganic Pigments}.
\newblock John Wiley \& Sons, 2008.

\bibitem{Kavan1996}
L.~Kavan, M.~Gr{\"{a}}tzel, S.~E. Gilbert, C.~Klemenz, and H.~J. Scheel,
  ``{Electrochemical and Photoelectrochemical Investigation of Single-Crystal
  Anatase},'' {\em J. Am. Chem. Soc.}, vol.~118, no.~28, pp.~6716--6723, 1996.

\bibitem{Khan2002}
S.~U.~M. Khan, M.~Al-Shahry, and W.~B. Ingler, ``{Efficient Photochemical Water
  Splitting by a Chemically Modified n-TiO2},'' {\em Science}, vol.~297,
  no.~5590, pp.~2243--2245, 2002.

\bibitem{Conesa2010}
J.~C. Conesa, ``{The Relevance of Dispersion Interactions for the Stability of
  Oxide Phases},'' {\em J. Phys. Chem. C}, vol.~114, no.~51, pp.~22718--22726,
  2010.

\bibitem{Arroyo2011}
M.~E. {Arroyo-de Dompablo}, A.~Morales-Garc{\'{i}}a, and M.~Taravillo, ``{DFT+U
  calculations of crystal lattice, electronic structure, and phase stability
  under pressure of TiO2 polymorphs},'' {\em J. Chem. Phys.}, vol.~135, no.~5,
  p.~54503, 2011.

\bibitem{Muscat2002}
J.~Muscat, V.~Swamy, and N.~M. Harrison, ``First-principles calculations of the
  phase stability of ${\mathrm{tio}}_{2}$,'' {\em Phys. Rev. B}, vol.~65,
  p.~224112, Jun 2002.

\bibitem{Labat2007}
F.~Labat, P.~Baranek, C.~Domain, C.~Minot, and C.~Adamo, ``{Density functional
  theory analysis of the structural and electronic properties of TiO2 rutile
  and anatase polytypes: Performances of different exchange-correlation
  functionals},'' {\em J. Chem. Phys.}, vol.~126, no.~15, p.~154703, 2007.

\bibitem{Hjorth2017}
A.~H. Larsen, J.~J. Mortensen, J.~Blomqvist, I.~E. Castelli, R.~Christensen,
  M.~Dułak, J.~Friis, M.~N. Groves, B.~Hammer, C.~Hargus, E.~D. Hermes, P.~C.
  Jennings, P.~B. Jensen, J.~Kermode, J.~R. Kitchin, E.~L. Kolsbjerg, J.~Kubal,
  K.~Kaasbjerg, S.~Lysgaard, J.~B. Maronsson, T.~Maxson, T.~Olsen, L.~Pastewka,
  A.~Peterson, C.~Rostgaard, J.~Schiøtz, O.~Schütt, M.~Strange, K.~S.
  Thygesen, T.~Vegge, L.~Vilhelmsen, M.~Walter, Z.~Zeng, and K.~W. Jacobsen,
  ``The atomic simulation environment—a python library for working with
  atoms,'' {\em J. Phys. Condens. Matter}, vol.~29, p.~273002, 2017.

\bibitem{Alchagirov2003}
A.~B. Alchagirov, J.~P. Perdew, J.~C. Boettger, R.~C. Albers, and C.~Fiolhais,
  ``Reply to ``comment on `energy and pressure versus volume: Equations of
  state motivated by the stabilized jellium model' '','' {\em Phys. Rev. B},
  vol.~67, p.~026103, Jan 2003.

\bibitem{Birch1947}
F.~Birch, ``Finite elastic strain of cubic crystals,'' {\em Phys. Rev.},
  vol.~71, pp.~809--824, Jun 1947.

\bibitem{Lee2015}
J.~Lee, D.-H. Seo, M.~Balasubramanian, N.~Twu, X.~Li, and G.~Ceder, ``A new
  class of high capacity cation-disordered oxides for rechargeable lithium
  batteries: Li–ni–ti–mo oxides,'' {\em Energy Environ. Sci.}, vol.~8,
  no.~11, pp.~3255--3265, 2015.

\end{thebibliography}

\end{document}